\documentclass[11pt,a4paper]{article}

\usepackage{geometry}
\usepackage{ulem}
\usepackage{tabularx}
\usepackage{booktabs}
\usepackage[colorlinks=true,linkcolor=blue,citecolor=blue,urlcolor=blue,bookmarks]{hyperref}
\usepackage{xcolor}

\newcommand{\affiliation}[1]{\\ \parbox[t]{0.9\textwidth}{\centering\small\textit{#1}}} 

\usepackage{amsmath,amssymb,bm,mathtools}
\usepackage{microtype}
\usepackage{graphicx}
\usepackage{tikz}
\usepackage[numbers,sort&compress]{natbib}

\newcommand{\AdS}{\mathrm{AdS}}
\newcommand{\dS}{\mathrm{dS}}
\newcommand{\Arg}{\operatorname{Arg}}
\newcommand{\dd}{\mathrm{d}}
\newcommand{\cC}{\mathcal{C}}

\newcommand{\Tr}{\operatorname{Tr}}

\newcommand{\Rea}{\operatorname{Re}}
\newcommand{\Ima}{\operatorname{Im}}

\usepackage{mathtools}
\mathtoolsset{showonlyrefs=true,showmanualtags=true}

\def \be {\begin{equation}}
\def \ee {\end{equation}}
\def \ba {\begin{array}}
\def \ea {\end{array}}
\def \bea{\begin{eqnarray}}
\def \eea{\end{eqnarray}}
\def \nn {\nonumber}

\usepackage{fancyhdr}

\fancypagestyle{plain}{
  \fancyhf{}
  \fancyhead[R]{\small YITP-26-96}
  
  \fancyfoot[C]{\thepage}}

\title{Selecting Complex Extremal Surfaces with the Kontsevich--Segal--Witten Criterion}
\author{Wu-zhong Guo$^{1,2,}$\footnote{wuzhong@hust.edu.cn}\\ \affiliation{$^1$School of Physics, Huazhong University of Science and Technology, \\
Luoyu Road 1037, Wuhan, Hubei 430074, China\\
$^2$Center for Gravitational Physics and Quantum Information,\\
 Yukawa Institute for Theoretical Physics, Kyoto University, Kyoto 606-8502, Japan}
}
\date{}

\begin{document}

\maketitle
\begin{abstract}
Complex extremal surfaces naturally arise in holographic observables associated with timelike subregions in AdS/CFT and with holographic observables in dS/CFT, but their integration contours are generally ambiguous. In this work, we propose a method for constructing complex bulk metrics from families of complex extremal surfaces, thereby generating candidate complex geometries relevant to the gravitational path integral. This construction provides a concrete realization of how complex bulk geometry may emerge from timelike entanglement, extending the familiar idea that spacetime geometry is encoded in quantum entanglement. We use the Kontsevich--Segal--Witten (KSW) criterion as a strong consistency condition to constrain the corresponding contours. The same framework also explains why spacelike entanglement naturally selects a real Lorentzian section. In several AdS and dS examples, the KSW condition uniquely determines the admissible contour within the class considered. We also identify configurations for which the resulting complex geometry violates the KSW bound near the asymptotic boundary, revealing both the scope and the limitations of the construction. These results highlight KSW admissibility as a useful organizing principle for complex saddles in real-time holography.

\end{abstract}



\tableofcontents
\section{Introduction}
The Ryu--Takayanagi (RT) and Hubeny--Rangamani--Takayanagi (HRT) prescriptions relate quantum entanglement to codimension-two extremal surfaces in a semiclassical bulk \cite{Ryu:2006bv,Hubeny:2007xt}. Building on this idea, substantial progress has reshaped our understanding of the information-theoretic structure of spacetime \cite{Swingle:2009bg,VanRaamsdonk:2010pw,Maldacena:2013xja,Almheiri:2014lwa,Jafferis:2015del,Dong:2016eik}. The quantum extremal surface proposal \cite{Engelhardt:2014gca} also provides a framework for understanding the information-loss problem \cite{Penington:2019npb,Almheiri:2019psf,Penington:2019kki,Almheiri:2019qdq,Almheiri:2020cfm}. A concise review of the development of this direction can be found in \cite{Takayanagi:2025ula}.

\begin{figure}[t]
\centering
\includegraphics[width=0.4\textwidth]{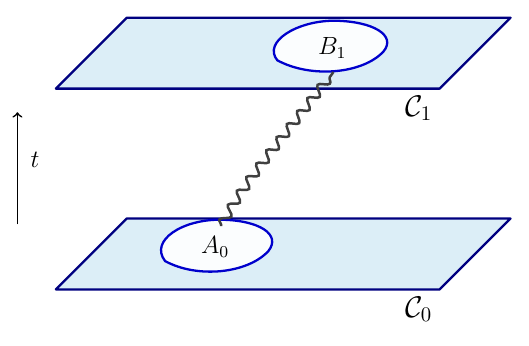}
\caption{Illustration of the reduced spacetime density matrix $T_{A_0B_1}$. The subsystems $A_0\subset\mathcal{C}_0$ and $B_1\subset\mathcal{C}_1$ lie on two different Cauchy surfaces, and the wavy line represents the temporal correlations encoded by $T_{A_0B_1}$.}
\label{fig:STDM}
\end{figure}

Extremal surfaces are generally associated with the entanglement or correlations of spatial subregions. In quantum field theories, however, timelike correlations are also significant. Such information is not manifest in the usual framework of spatial entanglement. In the quantum-information community, there has been a long-standing effort to formulate spatial and temporal correlations within a unified framework. Many different approaches have been developed, under different names and with an emphasis on different aspects \cite{Fitzsimons:2013gga,Buscemi:2013xlk,LS,Horsman:2017,Cotler:2017anu,Fullwood:2022rjd,Parzygnat:2022pax,Lie:2024kbl,Lie:2025wvb,Diaz:2020dfe,Milekhin:2025ycm,Diaz:2025aqe,Guo:2025dtq,Das:2025fcd,Diaz:2026qrf}; see \cite{Diaz:2026qrf} for a discussion of the relations among these different approaches. Recently, a similar idea has been applied to QFTs. By introducing a generally non-Hermitian operator that captures timelike correlations, it is possible to provide a well-defined definition and a Schwinger--Keldysh representation of entanglement in time \cite{Milekhin:2025ycm}. This provides an operator definition for understanding timelike entanglement and its holographic duality \cite{Doi:2022iyj,Heller:2024whi}. It was further shown in \cite{Guo:2025dtq} that one can first define the so-called spacetime density matrix for general Cauchy surfaces. Consider two Cauchy surfaces $C_0$ and $C_1$, associated with Hilbert spaces $\mathcal{H}_0$ and $\mathcal{H}_1$, respectively. The spacetime density matrix $T_{C_0C_1}:\mathcal{H}_0\otimes\mathcal{H}_1\to\mathcal{H}_0\otimes\mathcal{H}_1$ is defined by $\Tr(T_{C_0C_1}\mathcal{O}_0\otimes\mathcal{O}_1) = tr\bigl(\rho_0\,\mathcal{O}_0(t_0)\mathcal{O}_1(t_1)\bigr)$ 
for arbitrary operators $\mathcal{O}_0$ and $\mathcal{O}_1$ \cite{note1}. Given two subsystems $A_0$ and $B_1$ on $C_0$ and $C_1$, respectively, the reduced spacetime density matrix can be defined as $T_{A_0B_1}:=\Tr_{\bar{A}0\bar{B}1} T_{C_0C_1}$,
where $\bar{A}_0$ and $\bar{B}_1$ denote the complements of $A_0$ and $B_1$; see Fig.~\ref{fig:STDM} for an illustration. In general, $T_{A_0B_1}$ is non-Hermitian. The pseudo-R\	'enyi entropy can be defined as
\begin{equation}
S_n(T_{A_0B_1}):=\frac{\log\Tr(T_{A_0B_1}^n)}{1-n}.
\end{equation}
In this approach, the notions of spacelike and timelike entanglement are unified in a certain sense. It makes it possible to define entanglement for general subregions in spacetime, irrespective of their causal relation.

Timelike entanglement entropy is generally complex-valued, and its proposed bulk representatives are complex extremal surfaces, or equivalently, extremal surfaces evaluated on nontrivial cycles of a complexified geometry \cite{Heller:2024whi}. This provides a new probe of other aspects of holography, as well as of black-hole interiors \cite{Li:2022tsv,Doi:2023zaf,Narayan:2023ebn,Chu:2023zah,Das:2023yyl,Guo:2024lrr,Anegawa:2024kdj,Guo:2025pru,Guo:2025mwp,Nunez:2025ppd,Nunez:2025puk,Heller:2025kvp,Gong:2025pnu,Li:2025tud,Afrasiar:2025eam,Li:2026fcr,Katoch:2026dzs,Bernamonti:2026pxo}. On the other hand, in dS/CFT, holographic entanglement entropy is also expected to be dual to extremal surfaces \cite{Narayan:2015vda,Sato:2015tta,Doi:2022iyj,Narayan:2022afv,Nanda:2025tid}. Recently, it has been shown that complex extremal surfaces provide a more natural framework for general considerations \cite{Fujiki:2025rtx,Narayan:2026wzp}. This indicates that complex extremal surfaces may serve as bulk probes of non-Hermitian quantities on the field-theory side, which are closely related to causality or to the non-Hermitian nature of the theory \cite{Kawamoto:2025oko,Harper:2025lav}.
Let us briefly discuss why we expect complex extremal surfaces to be necessary. For a reduced spacetime density matrix $T_{A_0B_1}$, $S_n(T_{A_0B_1})$ can be evaluated using the real-time replica method \cite{Dong:2016hjy,Colin-Ellerin:2020mva,Colin-Ellerin:2021jev,Milekhin:2025ycm,Gong:2025pnu}. It admits a Lorentzian replica representation in which
\begin{equation}\Tr T_{A_0B_1}^n=Z[\mathcal{B}_n],\qquad
S_n(T_{A_0B_1})=\frac{1}{1-n}\log Z[\mathcal{B}_n],\label{eq:replica}
\end{equation}
where $\mathcal{B}_n$ is an $n$-fold replicated Lorentzian manifold, and $\mathcal{M}_n$ is a bulk geometry whose boundary is $\mathcal{B}_n$. Holographically, $Z[\mathcal{B}_n]$ is evaluated by a gravitational path integral with the oscillatory weight $e^{iI[g]}$. Even when the asymptotic boundary metric and the original integration variables are real, the corresponding steepest-descent cycles can therefore pass through complex saddles. A complete gravitational derivation should ultimately determine the relevant cycle from this Lorentzian replica path integral, plausibly through a Picard--Lefschetz analysis \cite{Witten:2010cx,Witten:2010zr,Harlow:2011ny}. Such an analysis must determine both the critical geometry and the intersection number of its thimble with the original integration contour. This situation is also familiar from real-time replicas for ordinary spatial entropy, where the Lorentzian saddle $\mathcal{M}_n$ is naturally described by complex or mixed-signature sections \cite{Colin-Ellerin:2020mva,Colin-Ellerin:2021jev,Held:2024qcl}. In the probe limit $n\to 1$, the HRT surface may lie on a real saddle, consistently with the Euclidean approach \cite{Lewkowycz:2013nqa}, and gives a real-valued entanglement entropy \cite{Dong:2016hjy}. For timelike entanglement entropy, however, the essential difference is that the entropy is generally complex-valued. We therefore expect the saddle to remain complex even in the probe limit. This is consistent with the proposal of \cite{Heller:2024whi}, in which the relevant bulk extremal surfaces live in a complexified geometry.

However, at present, it remains difficult to determine the complex saddle directly through a Picard--Lefschetz analysis. In this paper, we propose to construct possible saddles using complex extremal surfaces. The idea is to build the bulk geometry by collecting extremal surfaces associated with all possible subsystems in the boundary field theory. Specifically, consider a $d$-dimensional spacetime with metric $G_{MN}$
and coordinates $X^M$. Let $X^M=X^M(\lambda,\sigma^A;a^i)$ describe a continuous family of extremal
surfaces, where $a^i$ are real parameters specifying the boundary subsystem,
$\lambda$ and $\sigma^A$ denote the remaining real intrinsic coordinates on each extremal surface. Replacing $\lambda$ by a contour $\lambda=\lambda(u)$, with
$u\in\mathbb{R}$ \cite{note_parameter}, gives the pullback metric
\begin{equation}
g_{\alpha\beta}(q)=
G_{MN}(X)\,
\partial_\alpha X^M
\partial_\beta X^N,
\qquad
q^\alpha=(u,\sigma^A,a^i).
\label{eq:pullback}
\end{equation}
Wherever the family map is nondegenerate, the real coordinates $q^\alpha$
locally parametrize a real-dimensional cycle in the complexified bulk.
Thus, each contour image determines a complex family geometry, up to
orientation and regular real reparametrization. This construction is
analogous to the contour deformation of complex spacetime metrics discussed
by Witten \cite{Witten:2021nzp}.

The next question is how to select the appropriate saddle. Complex saddles arise as a general issue in gravitational path integrals \cite{Gibbons:1976ue,Gibbons:1978ac,Halliwell:1989dy,Hartle:1983ai,Louko:1995jw}. Witten proposed that allowable complex metrics can be selected using the pointwise condition of Kontsevich and Segal \cite{Kontsevich:2021dmb}, which ensures that the path integral is well defined for general real $p$-form field theories. This idea can be traced back to \cite{Louko:1995jw}, which may be viewed as a version of the scalar-field condition in $1+1$ dimensions. Specifically, when a complex metric is written in diagonal form in a real basis, $ds^2=\sum_\gamma\Lambda_\gamma(\dd q^\gamma)^2$, allowability requires
\begin{equation}
\Theta[g]\equiv\sum_\gamma\left|\operatorname{Arg}\Lambda_\gamma\right|<\pi.
\label{eq:KSW}
\end{equation}
Useful necessary conditions for a complex metric $g_{\alpha\beta}$, obtained from the scalar actions,  are
\begin{equation}
\operatorname{Re}\sqrt{g}>0,\qquad\operatorname{Re}\left(\sqrt{g}g^{\alpha\beta}\right)>0,
\label{eq:matrixksw}
\end{equation}
where the second inequality means that $\operatorname{Re}\left(\sqrt{g}g^{\alpha\beta}\right)$ is a positive matrix. These conditions guarantee the existence of a pointwise real basis in which the complex metric is diagonal. They do not, in general, guarantee
diagonalization by a real coordinate transformation in a neighborhood.
 In this paper, we also include the marginal cases in which equality holds in Eqs.~\eqref{eq:KSW} and \eqref{eq:matrixksw}, so that ordinary Lorentzian metrics are included in the allowable domain. A small sample of interesting applications of the KSW criterion can be found in \cite{Lehners:2021mah,Loges:2022nuw,Visser:2021ucg,Jonas:2022uqb,Hertog:2023vot,Chen:2023prz,Chen:2023sry,Ailiga:2025fny,Ailiga:2025osa,BenettiGenolini:2025jwe,BenettiGenolini:2026raa,Maldacena:2026jqd}. Our proposal is to apply the KSW criterion not to the intrinsic metric of a single extremal surface, but to the bulk cycle generated by a family of such surfaces. The KSW criterion then becomes a pointwise restriction on the contour $\lambda(u)$. We show below that it can select a unique contour in nontrivial examples and can also diagnose an obstruction. 

While previous applications have often used the KSW criterion to constrain
allowed parameter ranges, we find that, in several examples, it uniquely
selects the admissible contour within the class considered. This demonstrates
that the KSW criterion can serve as a remarkably strong selection principle
for the complex extremal surfaces.

\section{A universal complex metric in locally $\AdS_3$ backgrounds}
\begin{figure}[t]
\centering
\includegraphics[width=0.7\textwidth]{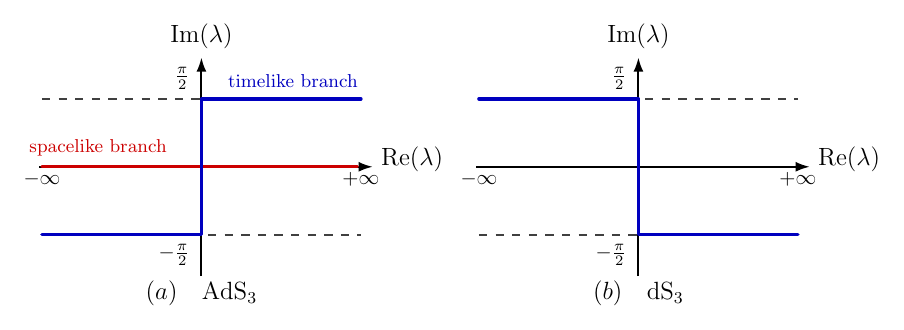}
\caption{
Integration contours in the complex $\lambda$ plane for
(a) AdS$_3$ and (b) dS$_3$. In (a), the red and blue curves correspond
to the spacelike and timelike branches of the KSW-selected contour,
respectively. In (b), the blue curve denotes the KSW-selected contour.
}
\label{fig:lambda}
\end{figure}

Consider first Poincar\'e $\AdS_3$ and geodesics anchored at $(t,x)=(\pm t_0,\pm x_0)$ with $t_0,x_0>0$. A unified parametrization for both spacelike and timelike EE is
\begin{equation}
 z=\frac{\sqrt{x_0^2-t_0^2}}{\cosh\lambda},\quad
 t=t_0\tanh\lambda,\quad x=x_0\tanh\lambda .
 \label{eq:ads3embed}
\end{equation}
The above parametrization applies to both the spacelike and timelike
branches. For the spacelike branch, $t_0<x_0$, the two endpoints are
reached as $\lambda\to-\infty$ and $\lambda\to+\infty$, respectively. For the timelike branch, $t_0>x_0$, the two endpoints correspond to
$\lambda\to-\infty-i\pi/2$ and
$\lambda\to+\infty+i\pi/2$, respectively. See Fig.\ref{fig:lambda} for an illustration.

With the new real coordinates $t_0=\rho_t\cosh\eta, x_0=\rho_t\sinh\eta,\xi=\log\rho_t$, there is one-to-one correspondence between every curve $\lambda(u)$  and the following complex metric 
\begin{equation}
 \begin{aligned}
 \frac{ds^2_\mathbb{C}}{L_{\text{AdS}}^2}=\dd \lambda^2
 +\cosh^2\lambda\dd\xi^2-\sinh^2\lambda\dd\eta^2,
 \end{aligned}
 \label{eq:ads3metric}
\end{equation}
where $L_{\text{AdS}}$ is AdS radius.
The important point is that Eq.~\eqref{eq:ads3metric} is not special to the Poincar\'e patch. Direct constructions from global-$\AdS_3$ geodesics and from nonrotating BTZ geodesics reduce to precisely the same metric. See the Appendix for the details. The quotient defining BTZ only changes global identifications and coordinate ranges. Hence Poincar\'e $\AdS_3$, global $\AdS_3$, and BTZ define one and the same local complex KSW problem, rather than three analogous examples.

Write $\lambda=g+if$, with $g,f\in\mathbb R$, take $g$ piecewise monotonic, and restrict to $-\pi/2\leq f\leq\pi/2$. The three absolute phases in Eq.~\eqref{eq:ads3metric} are
\begin{align}
 \phi_u&=2\arctan\left|\frac{f'}{g'}\right|,\nonumber\\
 \phi_\xi&=2\arctan|\tanh g\tan f|,\nonumber\\
 \phi_\eta&=\pi-2\arctan|\coth g\tan f|.
 \label{eq:ads3phases}
\end{align}
The branch choices are continuous from the Lorentzian endpoint sections. KSW condition including the equality case gives \cite{note_marginal}
\begin{equation}
 \left|\frac{f'}{g'}\right|
 \leq \left|\frac{\sin 2f}{\sinh 2g}\right| .
 \label{eq:diffineq}
\end{equation}
Defining
\begin{equation}
 Q_{\text{AdS}}=\frac{|\tanh g|}{|\tan f|^{\sigma}},\qquad
 \sigma=\operatorname{sgn}(gg')\operatorname{sgn}(ff'),
\end{equation}
one finds that $Q_{\text{AdS}}$ is monotonic in the regions what we are considering. 
The endpoint values and the monotonicity of $Q_{\text{AdS}}$ would give a strong constraint on the shape of the curve $\lambda$. For the spacelike EE, the only possible curve is the one with $f=0$, that is $\lambda$ is real. While for timelike EE, the curve should take the following one:
\begin{equation}
 \cC_{\AdS}:\quad
 \lambda=\begin{cases}
 g-i\pi/2,&g<0,\\
 if,&g=0,\quad f\in (-\frac{\pi}{2},\frac{\pi}{2}),\\
 g+i\pi/2,&g>0.
 \end{cases}
 \label{eq:threepiece}
\end{equation}
See Fig.\ref{fig:lambda} for an illustration of the curve.
The abolve statement holds simultaneously in the Poincar\'e, global, and BTZ realizations. One could refer to the Appendix for the details of the derivation. Note that the contour selected here coincides precisely with that used in \cite{Heller:2024whi,Guo:2025pru}.

The spacelike EE would give the Lorentzian AdS$_3$, which is consistent with the expectation. For timelike EE, the metric is new, each patch corresponding to Eq.\eqref{eq:threepiece} saturates the marginal case of KSW. One could check directly the metric is also Lorentzian,
\begin{equation}
 \frac{ds^2_{\text{hor}}}{L_{\text{AdS}}^2}=\dd g^2-\sinh^2g\,\dd\xi^2
 +\cosh^2g\,\dd\eta^2 \quad \text{with}\quad \lambda=g\pm i\frac{\pi}{2},
 \label{eq:patchmetrics:hor}
\end{equation}
\begin{equation}
 \frac{ds^2_{\text{mid}}}{L_{\text{AdS}}^2}=-\dd f^2+\cos^2f\,\dd\xi^2
 +\sin^2f\,\dd\eta^2 \quad \text{with}\quad \lambda=i f.
 \label{eq:middle}
\end{equation}
Thus all three locally-$\AdS_3$ backgrounds select the same complex cycle, assembled from different Lorentzian real sections of the same complex metric $ds^2_\mathbb{C}$.

\section{Universal Lorentzian sector}

On the middle segment of the KSW-selected contour, $\lambda=if$, this
metric becomes \eqref{eq:middle}.
This is a real Lorentzian patch of pure AdS$_3$, which can be identified
with the interior of an AdS-Rindler wedge by setting $r=\sin f$ in the
region $0<r<1$. Thus, the middle part of the contour is not an
intrinsically complex local spacetime, but another real Lorentzian
section of the complexified AdS$_3$ geometry.

For both Poincar'e AdS$_3$ and the BTZ black hole, the KSW criterion selects the same three-piece contour in the complex affine-parameter plane, and the corresponding extremal-surface-adapted metric takes the same local AdS--Rindler form. The selected contour therefore decomposes the complexified geometry into two exterior-type real Lorentzian sections connected by an AdS--Rindler interior section. The two examples differ, however, in their global embedding. In the Poincar'e case, the exterior branches extend directly from the asymptotic boundary to the Poincar'e horizons. In the BTZ case, each branch crosses the BTZ event horizon and reaches the $r=0$ quotient singularity before continuing into an additional real section of the complexified BTZ geometry and eventually reaching the adapted AdS--Rindler horizon. Thus, the Poincar'e and BTZ constructions share the same local AdS--Rindler structure, while their global realizations differ because of the BTZ quotient. An illustration of the metric is shown in Fig.\ref{fig:ads_ds_geometry} for the Poincar\'e case. Further details are provided in the appendix. 

This construction is analogous to the familiar Euclidean--Lorentzian geometries that arise in state preparation and real-time path integrals. In both cases, different real sections are connected by analytic continuation within a common complexified geometry. The essential distinction is that, in the present construction, all open sections are Lorentzian: the analytic continuation changes the underlying real structure rather than the metric signature. Accordingly, the resulting geometry should be understood as a piecewise-Lorentzian complex saddle, rather than as a real Lorentzian spacetime obtained by gluing along a null hypersurface. It is also interesting to note that a geometry involving two AdS$_3$ patches also appears in the proposal for holographic entanglement entropy in $T\bar T$ deformed CFTs with a positive deformation parameter \cite{Apolo:2023vnm,Apolo:2023ckr}, although it arises from a completely different construction.
 
For fixed values of the family coordinates $\xi$ and $\eta$, the extremal surface becomes timelike along the middle segment of the KSW-selected contour. This provides a geometric origin for the imaginary contribution to timelike entanglement entropy and is closely related to the original proposal in which the imaginary part is attributed to a timelike extremal segment \cite{Doi:2022iyj}. The essential difference is that, in the present construction, this timelike segment does not lie in the original Lorentzian section of the bulk. Instead, it arises as part of a single complex extremal surface when the contour passes through a different real Lorentzian section of the complexified geometry.


\section{The $\dS_3$ counterpart}
A closely parallel construction applies to global $\dS_3$,
\begin{equation}
 \frac{\dd s^2}{L_{\mathrm{dS}}^2}
 =
 -\dd T^2
 +\cosh^2 T
 \left(
 \dd\theta^2+\sin^2\theta\,\dd\varphi^2
 \right),
 \label{eq:ds_global}
\end{equation}
where $L_{\mathrm{dS}}$ denotes the radius of the dS spacetime.
We consider complex extremal curves anchored at the two points $
 (\theta,\varphi)=
 \left(
 \frac{\pi}{2}-\theta_0,\pi-\varphi_0
 \right),
 \left(
 \frac{\pi}{2}+\theta_0,\pi+\varphi_0
 \right)$ 
on future infinity $\mathcal I^+$. Defining
\begin{equation}
 \Delta
 :=
 \sin^2\theta_0
 +\cos^2\theta_0\sin^2\varphi_0,
\end{equation}
the extremal curves can be parametrized as
\begin{align}
 T
 &=
 \operatorname{arcsinh}
 \left(
 \frac{i\cosh\lambda}{\sqrt{\Delta}}
 \right),
 \nonumber\\
 \theta
 &=
 \arccos
 \left(
 \frac{\sin\theta_0\sinh\lambda}
 {\sqrt{\cosh^2\lambda-\Delta}}
 \right),
 \nonumber\\
 \varphi
 &=
 \pi+
 \arctan
 \left(
 -\tan\varphi_0\tanh\lambda
 \right).
 \label{eq:ds_extremal_curve}
\end{align}
The two asymptotic endpoints correspond to
 $\lambda\longrightarrow \pm\infty\mp\frac{i\pi}{2}.$

After a real redefinition of the family parameters
$(\theta_0,\varphi_0)\to(\chi,\eta)$, a contour
$\lambda=\lambda(u)$ induces the diagonal metric
\begin{equation}
 \frac{\dd s_{\mathbb C}^2}{L_{\mathrm{dS}}^2}
 =
 -\lambda'(u)^2\dd u^2
 -\cosh^2\lambda(u)\,\dd\chi^2
 -\sinh^2\lambda(u)\,\dd\eta^2.
 \label{eq:ds_family_metric}
\end{equation}
This is the dS counterpart of the extremal-surface-adapted metric
encountered in the $\AdS_3$ examples.  The KSW criterion implies
\begin{equation}
\left|\frac{f'}{g'}\right|
\geq
\left|\frac{\sin\bigl(2f\bigr)}
{\sinh\bigl(2g\bigr)}\right|,
\label{eq:inequality:ds}
\end{equation}
which is precisely the reverse of the corresponding inequality in the
AdS$_3$ case.

As in the $\AdS_3$ analysis, the KSW condition can be expressed through
a non-negative monotonic quantity. For dS$_3$ case a convenient choice is
\begin{equation}
Q_{\mathrm{dS}}(u)
:=
\frac{|\tanh g(u)|}{|\tan f(u)|^{\sigma}},
\qquad
\sigma
:=
\operatorname{sgn}\bigl(gg'\bigr)
\operatorname{sgn}\bigl(ff'\bigr),
\label{eq:ds_monotonic_quantity}
\end{equation}
where $\lambda(u)=g(u)+if(u)$ with $g(u),f(u)\in\mathbb R$. On each
regular branch with fixed signs of $g$, $g'$, $f$, and $f'$, the sign
factor $\sigma$ is constant, and the full KSW phase condition implies a
definite monotonicity of $Q_{\mathrm{dS}}$.
Together with the endpoint condition, this excludes every
deformation away from the horizontal branches. Within the same regular,
piecewise-monotonic class used in the $\AdS_3$ analysis, the contour is
therefore uniquely fixed to be
\begin{equation}
 \mathcal C_{\mathrm{dS}}:\qquad
 \lambda=
 \begin{cases}
  g+i\pi/2,
  & g<0,
  \\[2mm]
  if,
  & g=0, \, f\in (-\frac{\pi}{2},\frac{\pi}{2})
  \\[2mm]
  g-i\pi/2,
  & g>0.
 \end{cases}
 \label{eq:ds_selected_contour}
\end{equation}
The detailed monotonicity and uniqueness proof is given in
Appendix. The contour found here is identical to that used in \cite{Fujiki:2025rtx}.

\begin{figure*}[t]
 \centering
 \begin{minipage}[t]{0.48\textwidth}
  \centering
  \includegraphics[width=\linewidth]{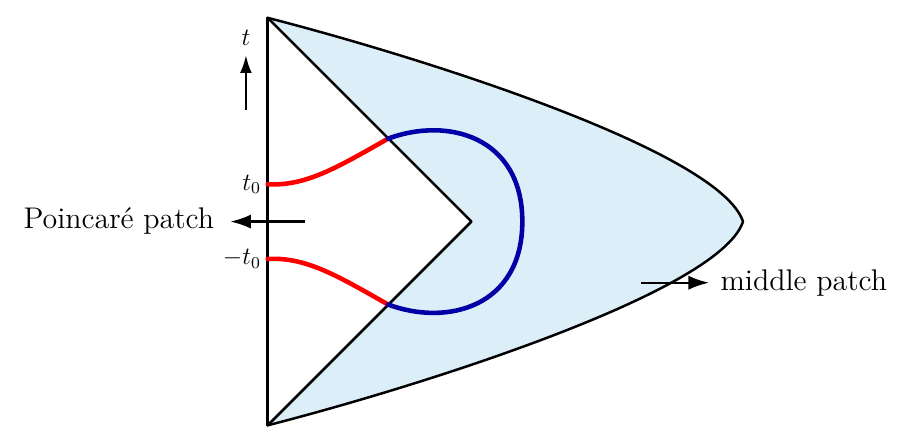}\\[-0.5ex]
  {\small (a) $\AdS_3$}
 \end{minipage}
 \hfill
 \begin{minipage}[t]{0.40\textwidth}
  \centering
  \includegraphics[width=\linewidth]{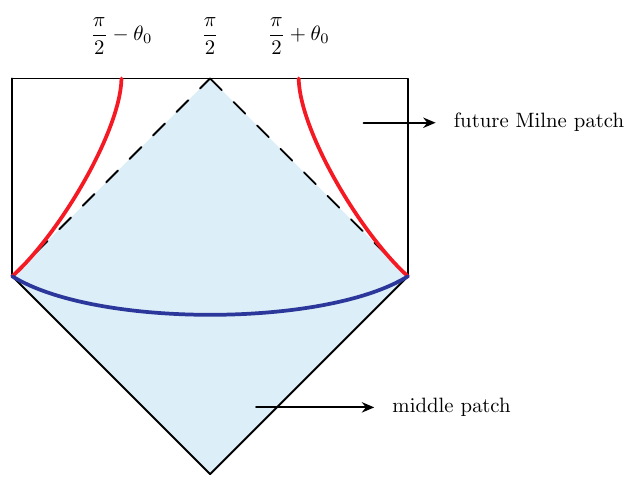}\\[-0.5ex]
  {\small (b) $\dS_3$}
 \end{minipage}
 \caption{
Schematic representations of the geometries associated with the
KSW-selected complex extremal curves in (a) Poincar\'e AdS$_3$ and
(b) global dS$_3$.
In the AdS$_3$ case, the geometry consists of the original Poincar\'e
patch and a middle Lorentzian patch reached across the Poincar\'e
horizons. The red outer portions of the extremal curve are spacelike,
whereas the blue portion in the middle patch is timelike.
In the dS$_3$ case, the geometry consists of the original future
cosmological section and a middle Lorentzian section locally
isometric to static patch. These sections admit a smooth continuation across the corresponding cosmological Killing horizons associated with the static observer at $\theta=\frac{\pi}{2}$, while the KSW-selected extremal curve itself passes through
their bifurcation loci at $T=0$ in the global coordinate. Relative to the AdS$_3$ case, the causal
characters are interchanged: the red outer portions are timelike,
whereas the blue segment in the middle patch is spacelike.
 }
 \label{fig:ads_ds_geometry}
\end{figure*}

As in the Poincar'e AdS$_3$ example, the KSW-selected contour
$\mathcal C_{\mathrm{dS}}$ decomposes into several real Lorentzian
sections of the complexified geometry, connected across horizon-like
surfaces, as illustrated in Fig.~\ref{fig:ads_ds_geometry}.
The two horizontal branches correspond to the original cosmological
section of dS$_3$, while the middle vertical branch gives another real
Lorentzian section. The detailed geometric interpretation and the
verification that each segment saturates the KSW bound are presented in
Appendix. Our approach should extend naturally to higher-dimensional de Sitter
spacetimes. Holographic entanglement entropy in higher-dimensional
de Sitter spacetimes has been studied extensively, see, e.g., \cite{Narayan:2015vda,Narayan:2022afv,Anastasiou:2025rvz,Anastasiou:2026bbf}. It would be interesting to determine how the KSW criterion
constrains the corresponding bulk geometry, especially since the
geometry obtained in our $\dS_3$ example differs substantially from the
usual Hartle--Hawking construction. This may provide a new perspective
on de Sitter holography. Finite-$n$ replica geometries in $\dS_3$ and $\dS_4$ have
recently been constructed \cite{Nanda:2025tid}, and testing their
KSW allowability would provide an important extension beyond the
$n\to1$ limit considered here.

\section{Higher-dimensional hyperbolic surfaces}
A direct higher-dimensional analogue arises in Poincar\'e $\AdS_4$. Let
$ \Delta=t_0^2-x_0^2$, the boundary entangling curve is chosen to be 
\begin{equation}
E_{t_0,x_0}:\qquad
t_0x-x_0t=0,
\qquad
t^2-x^2-y^2=\Delta.
\end{equation}
For $\Delta<0$, the entangling curve is a spatial circle lying on a boosted Cauchy slice, whereas for $\Delta>0$, it becomes a future–past pair of hyperbolic branches forming a timelike-separated boundary configuration. In both cases, the corresponding bulk extremal surface can be derived from the extremal-area equation and expressed in the unified quadratic form:
\begin{equation}
 t_0x-x_0t=0,\
 t^2-x^2-y^2-z^2=\Delta .
 \label{eq:quadricsurface}
\end{equation}
For $\Delta<0$, it is the ordinary hemispherical RT surface in a boosted Cauchy slice. For $\Delta>0$, it is a timelike hyperbolic surface. Writing $t_0=\rho_t\cosh\beta$, $x_0=\rho_t\sinh\beta$, a convenient parametrization is
\begin{align}
 t&=t_0\cosh\lambda\cosh\eta,&
 x&=x_0\cosh\lambda\cosh\eta,\nonumber\\
 y&=\rho_t\cosh\lambda\sinh\eta,&
 z&=\rho_t\sinh\lambda .
 \label{eq:hyperembed}
\end{align}
The future and past boundary branches occur at $\lambda=0$ and $\lambda=i\pi$. With $\xi=\log\rho_t$ and $\dd H_2^2=\dd\eta^2+\cosh^2\eta\,\dd\beta^2$, the family metric is
\begin{equation}
 \frac{ds^2}{L_{\text{AdS}}^2}=\frac{\dd\lambda^2-\dd\xi^2}{\sinh^2\lambda}
 +\coth^2\lambda\,\dd H_2^2.
 \label{eq:hypermetric1}
\end{equation}
Now define $\chi=\log\tanh\frac{\lambda}{2}-\frac{i\pi}{2}$.
Equation~\eqref{eq:hypermetric1} becomes exactly
\begin{equation}
 \frac{ds^2}{L_{\text{AdS}}^2}=\dd\chi^2+\cosh^2\chi\,\dd\xi^2
 -\sinh^2\chi\,\dd H_2^2.
 \label{eq:hypermetric2}
\end{equation}
The geometry constructed here closely resembles that arising in the Casini-Huerta-Myers approach to holographic entanglement entropy for spherical regions \cite{Casini:2011kv}, and this connection deserves further investigation. 
The regulated endpoints map to $\chi\to-\infty-i\pi/2$ and $\chi\to+\infty+i\pi/2$, precisely as the AdS$_3$ case.

For Eq.~\eqref{eq:hypermetric2}, let $\phi_u$, $\phi_\xi$, and $\phi_H$ denote the phases of $\chi'^2$, $\cosh^2\chi$, and $-\sinh^2\chi$. The full $\AdS_4$ bound is
\begin{equation}
 \Theta_4=\phi_u+\phi_\xi+2\phi_H\leq\pi.
\end{equation}
It implies the $\AdS_3$ inequality $\phi_u+\phi_\xi+\phi_H\leq\pi$. Hence the monotonicity theorem first restricts the contour to Eq.~\eqref{eq:threepiece} with $\lambda$ replaced by $\chi$. Direct substitution then proves sufficiency: on the horizontal pieces $(\phi_u,\phi_\xi,\phi_H)=(0,\pi,0)$, while on the vertical piece it is $(\pi,0,0)$. Therefore $\Theta_4=\pi$ pointwise. The extra hyperbolic directions carry zero phase and do not spoil allowability.

The spacelike sector is qualitatively different from the timelike one.
In an appropriate boosted frame, the boundary subsystem is a spatial
disk, and the corresponding extremal surface is the ordinary
hemispherical RT surface. In this case, the real contour $0<\lambda\leq \frac{\pi}{2}$ satisfies and saturates the KSW bound. Moreover, regularity at a positive
real radial cutoff excludes nontrivial complex deformations connected to
this branch. Thus, while the timelike sector is described by the
KSW-selected three-piece complex contour, the spacelike sector remains
entirely within the standard Lorentzian AdS section. The detailed KSW
analysis is given in Appendix.

The timelike hyperbola and the spacelike circle are therefore different
real sections of the same complexified quadratic family, with the KSW
criterion selecting the appropriate section in each causal sector.


\section{Strips and a KSW obstruction}
The construction can also be generalized to strip subregions in higher-dimensional AdS$_{d+1}$. For spacelike strips, it always reproduces a real Lorentzian AdS geometry, independent of the spacetime dimension. For timelike strips, however, the situation is qualitatively different. In particular, the near-boundary structure exhibits a distinct dependence on the parity of the boundary dimension $d$, leading to different realizations of the KSW criterion in even and odd dimensions.

For a planar strip in $\AdS_{d+1}$, the extremal surface can be written
\begin{equation}
 t=\frac{t_0}{c_d}\lambda,\quad
 x=\frac{x_0}{c_d}\lambda,\quad
 z=\frac{\sqrt{x_0^2-t_0^2}}{c_d}\,\zeta(\lambda),
 \label{eq:stripeq}
\end{equation}
where
\begin{equation}
 c_d-\lambda=\int_0^{\zeta}\frac{s^{d-1}\dd s}
 {\sqrt{1-s^{2(d-1)}}}
\end{equation}
near one endpoint. For timelike separation, $\sqrt{x_0^2-t_0^2}=i\rho_t$. We now employ a perturbative analysis. The leading near-boundary behavior is fixed to be
\begin{equation}
\zeta(u)=-iu+O(u^2),
\qquad
\lambda(u)=c_d-\frac{(-i)^d}{d}u^d+O(u^{d+1}).
\label{eq:generalphase}
\end{equation}
This boundary phase is not a freely chosen deformation of the contour. Rather, it follows from the extremal-surface equation together with the requirement of a positive real radial cutoff. In the Appendix, we further show that the same leading-order phase is independently implied by the KSW criterion.

In $\AdS_4$ (\(d=3\)), the complex metric in the coordinates
$(u,\rho_t,\eta,y)$, with
$t_0=\rho_t\cosh\eta$ and
$x_0=\rho_t\sinh\eta$,
can be constructed from Eq.~(\ref{eq:stripeq}) as
\begin{align}
&\frac{ds^2}{L^2}
=
\frac{\zeta'^2}{\zeta^2(1-\zeta^4)}\,du^2
+\frac{\lambda^2+\zeta^2}{\zeta^2}
\frac{d\rho_t^2}{\rho_t^2}
\nonumber\\
&
+\frac{2\zeta'}{\rho_t}
\left(
\frac1\zeta
-\frac{\lambda}{\sqrt{1-\zeta^4}}
\right)
du\,d\rho_t
-\frac{\lambda^2}{\zeta^2}\,d\eta^2
-\frac{c_3^{\,2}}{\rho_t^{\,2}\zeta^2}\,dy^2 .
\label{eq:timelikestripmetric}
\end{align}

The coordinates adapted to the family contain a non-diagonal $(u,\rho_t)$ block, so we first impose Eq.~\eqref{eq:matrixksw} rather than assuming a diagonal coordinate system. Expanding the most general local contour,
\begin{equation}
 \lambda=c_3-\frac{i}{3}u^3+\alpha_1u^4+\alpha_2u^5+\cdots,
\end{equation}
KSW positivity \eqref{eq:matrixksw} at the first two nontrivial orders imposes 
 $\operatorname{Re}\alpha_1=c_3$ and $ \operatorname{Re}\alpha_2=-4c_3\operatorname{Im}\alpha_1$. The constraints become mutually incompatible at the order involving the third subleading contour coefficient. Therefore, no contour within the general perturbative family can satisfy the KSW
condition \eqref{eq:matrixksw} in a punctured neighborhood of the boundary. 

As a complementary check, we also consider a restricted class for which
the metric can be locally diagonalized by a real coordinate
transformation. In the resulting diagonal frame,
\begin{equation}
\Theta[g]=\pi+\tilde{c} u+O(u^2), \qquad
\tilde c>0,
\label{eq:violation}
\end{equation}
where $z\propto u$ as $u\to0$. Thus the KSW phase sum exceeds $\pi$ already at order $O(z)$. Both analyses therefore show that the timelike-strip geometry violates the KSW criterion near the
conformal boundary. The spacelike strip behaves oppositely: the same analysis forces the contour coefficients to be real and reproduces Lorentzian AdS$_4$.

Equation~\eqref{eq:generalphase} also exposes a parity-dependent analytic structure. Along the ray $\zeta=-iq$, one has
\begin{equation}
c_d-\lambda
=
\int_0^{-iq}
\frac{s^{d-1}\,\mathrm{d}s}
{\sqrt{1-s^{2(d-1)}}}.
\label{eq:rayintegral}
\end{equation}
For even $d$, the right-hand side is real on the branch continuously connected to the origin, and the corresponding family metric admits a real Lorentzian presentation. For odd $d$, it is purely imaginary before the first branch point, so the embedding is locally complex. The branch points are
$\zeta_k=e^{i\pi k/(d-1)}$, with
$k=0,\ldots,2d-3$, and the negative imaginary ray reaches a branch point precisely when $d$ is odd. However, the two boundary branches do not in general meet at this point, so reaching the branch point does not by itself produce a connected timelike extremal surface.
The distinction is explicit in $\AdS_5$, for which $d=4$. The straight ray
\begin{equation}
\zeta=-iu,
\qquad
\lambda
=
c_4-\int_0^u
\frac{v^3\,\mathrm{d}v}{\sqrt{1+v^6}}
\label{eq:ads5ray}
\end{equation}
induces a real Lorentzian metric and saturates the KSW bound. Moreover, no branch point lies on the negative imaginary axis, since
$1-(-iu)^6=1+u^6>0$, and the ray can be extended toward the Poincaré horizon. It does not, however, return to the second boundary component. A complete timelike extremal surface must therefore eventually leave this real branch and follow a genuinely complex continuation, whose interior KSW behavior remains to be determined. Conversely, for odd $d$, the real-cutoff branch is locally complex and, in the $\AdS_4$ example studied below, violates the KSW condition. This does not by itself establish a universal KSW no-go theorem in every odd dimension. The local boundary phase, global sheet connectivity, and KSW allowability remain logically distinct.

The contrast between the strip and hyperbolic examples shows that the
KSW behavior is not determined solely by the bulk dimension or by the
timelike nature of the boundary separation; it also depends on the
chosen family of extremal surfaces. For spacelike subsystems, different
regular families are expected to provide different adapted
coordinatizations of the same Lorentzian AdS section. For timelike
subsystems, by contrast, no canonical real Lorentzian foliation exists,
and different subsystem geometries can lead to genuinely different
real-dimensional sections of the complexified bulk. This is illustrated
by the timelike strip, whose induced metric violates the KSW condition,
and the hyperbolic family, for which KSW selects a three-piece contour
composed of real Lorentzian sections. This configuration dependence may
encode part of the problem of selecting the appropriate complex saddle
and integration cycle in the Lorentzian gravitational path integral,
and deserves further investigation.



\section{Discussion} In this work, we proposed a direct relation between complex extremal surfaces and complex bulk metrics. By complexifying the extremal-surface parameter and applying the KSW criterion, we identified a restricted class of candidate complex saddles, see Table.\ref{table_I} for the summary. Our results suggest a complex extension of the familiar statement that spacetime emerges from entanglement: \textit{complex geometry emerges from timelike entanglement}. This is also consistent with the appearance of non-Hermitian operators and complex entanglement quantities in the boundary theory, which are naturally expected to be associated with complex saddles in the gravitational path integral \cite{Arean:2019pom,Arean:2024gks,Xian:2023zgu,Maeda:2026awj}. Our construction, however, provides only a possible selection principle and is not yet a first-principles derivation of the relevant integration cycle. It would be interesting to investigate whether the new geometry constructed here can also be used to evaluate correlation functions at timelike separation and how it is related to existing prescriptions in real-time holography \cite{Son:2002sd,Herzog:2002pc,Skenderis:2008dh,Skenderis:2008dg}.

For timelike strips in $\text{AdS}_{4}$, the present ansatz does not yield a KSW-allowable saddle. This may reflect the restricted nature of the construction, since only one parameter of a higher-dimensional extremal surface is complexified, or it may indicate that the relevant saddle genuinely lies outside the standard KSW domain. In $\text{AdS}_4$, the resulting metric also provides a complex generalization of the Fefferman--Graham expansion \cite{Fefferman:1985aa,Henningson:1998gx,deHaro:2000vlm}, whose holographic meaning remains to be understood. Moreover, the same framework can be applied to $\text{dS}$, and the physical interpretation of the KSW-selected geometries in both $\text{AdS}$ and $\text{dS}$ deserves further investigation.

\begin{table}[t]
\centering
\caption{Action of KSW on the principal extremal-surface families.}
\label{tab:summary}
\footnotesize
\begin{tabular*}{\textwidth}{@{\extracolsep{\fill}}ll}
\hline
Family & KSW-selected cycle \\
\hline
Poincar\'e/global/BTZ $\text{AdS}_3$ & real (space), $\mathcal{C}_{\text{AdS}}$ (time) \\
$\text{dS}_3$ & $\mathcal{C}_{\text{dS}}$ (time) \\
$\text{AdS}_4$ hyperbolic/ball & $\mathcal{C}_{\text{AdS}}$ / real hemisphere \\
$\text{AdS}_4$ strip & real (space), violation (time) \\
\hline
\end{tabular*}
\label{table_I}
\end{table}

~\\~\\~\\~\\~\\~\\
\textbf{Acknowledgments}\\

The author is grateful to  Ignacio J. Araya, Avijit Das, Kosei Fujiki, Chao-Qiang Geng, Peng-Xiang Hao, Michal P. Heller, Liang Li, Jian-Xin Lu, Chen-Te Ma, Javier Moreno, K. Narayan, Shan-Ming Ruan,  Kotaro Shinmyo, Jia-Rui Sun, Yu-ki Suzuki, Tadashi Takayanagi and Hongbao Zhang for valuable discussions. The author also thanks the organizers of the ``Holographic Universe 2026'' and ``Holography and Quantum Information'' workshop at the Yukawa Institute for Theoretical Physics (YITP), Kyoto University, and the ``2026 Workshop on Frontiers of Gravitation, Cosmology and Particle Physics'' at the University of Science and Technology of China for providing stimulating research environments. The author also thanks ChatGPT (OpenAI, GPT-5.6 and GPT-5.5) and Gemini
(Google, Gemini 3.5) for exploratory discussions, preliminary
algebraic checks, and consistency checks of some derivations,
specifically the higher-dimensional analyses, as well as
for language and \LaTeX{} assistance. 
 
This work was supported by the Seventh Young Faculty Development Program of Huazhong University of Science and Technology and by the Hubei Provincial Natural Science Foundation of China under Grant No.~2025AFB557.

\appendix

\section{Basic idea}
This appendix supplies detailed derivations for all results used in the main text. We retain the intermediate coordinate transformations, pullback metrics, phase identities, monotonicity arguments, near-boundary series, and matrix-positivity tests that are useful for independently checking the conclusions. We first formulate KSW allowability on a real-dimensional cycle. We then show explicitly that Poincar\'e $\AdS_3$, global $\AdS_3$, and nonrotating BTZ induce the same complex family metric and prove their common contour-selection theorem. A separate section derives the $\dS_3$ family metric and proves the reflected contour theorem from the full KSW phase cone and the reversed endpoint conditions. We next give the complete hyperbolic construction in $\AdS_4$, followed by the general-dimensional planar-strip equations, the local $\AdS_5$ branch, the full near-boundary analysis of the $\AdS_4$ strip, and an $\AdS_3$ perturbative consistency check.

The basic idea is to use a continuous family of extremal surfaces as a coordinate system for the bulk geometry. For each boundary subregion, the corresponding extremal surface is labeled by a set of boundary parameters, while the position along the surface is described by additional real parameters. Taken together, these variables define a real-dimensional slice of the complexified bulk spacetime. When the extremal surfaces are ordinary real surfaces, this construction is simply a change of coordinates in the original Lorentzian geometry. For timelike subregions, however, the relevant extremal surfaces generally extend into the complexified bulk. By allowing the surface parameter to follow a contour in a complex parameter plane, while keeping the labels of the surface family and the contour parameter real, one obtains a complex metric as the pullback of the complexified bulk metric onto this real-dimensional slice. In this way, a choice of complex extremal-surface contour is promoted to a choice of complex bulk geometry. The KSW criterion can then be applied pointwise to the resulting metric, providing a geometric constraint on which complex contours, and hence which complex extremal surfaces, may be regarded as admissible.

\section{KSW allowability on a real integration cycle}
\label{sec:ksw}

Let $\mathcal M_{\mathbb C}$ be a complexified spacetime and let $\mathcal C\subset\mathcal M_{\mathbb C}$ be a real-dimensional cycle parametrized by real coordinates $q^\alpha$. The pullback of the complex bulk metric is a complex symmetric bilinear form on the real tangent bundle $T_{\mathcal{C}}$,
\begin{equation}
 g_{\alpha \beta}(q)=G_{MN}(X(q))\,\partial_\alpha X^M\partial_\beta X^N.
 \label{S:eq:pullback}
\end{equation}
The relevant notion of equivalence is a real change of basis in $T_{\mathcal{C}}$. A holomorphic coordinate transformation that does not preserve the real cycle is not an innocuous diagonalization: it rotates the integration contour of the matter fields and may change convergence.

At a point where the quadratic form can be diagonalized by a real basis,
\begin{equation}
 g=\sum_{\gamma=1}^{D}\Lambda_\gamma (e^\gamma)^2,
 \qquad e^\gamma\in T^*_{\mathcal{C}}\ \text{real},\nn
\end{equation}
KSW allowability is
\begin{equation}
 \Theta[g]:=\sum_{\gamma=1}^{D}|\Arg\Lambda_\gamma|<\pi.
 \label{S:eq:ksw}
\end{equation}
We use the closure $\Theta\leq\pi$ in order to include ordinary Lorentzian metrics as the zero-regulator limit. The strict inequality is recovered by the usual $i\epsilon$ displacement.

For a non-diagonal metric, useful necessary conditions follow from the convergence of the scalar kinetic term and of the top-form term. With a continuous choice of $\sqrt{g}$, they are
\begin{equation}
 \Rea\sqrt{g}>0,
 \qquad
 M^{\alpha \beta}:=\Rea\!\left(\sqrt{g}\,g^{\alpha \beta}\right)> 0,
 \label{S:eq:matrixnecessary}
\end{equation}
where the second inequality means positive semidefiniteness as a real matrix. Again we will include the equality case in our discussions. When an explicit real diagonal frame is available, Eq.~\eqref{S:eq:ksw} is the complete condition and will be used directly. For the strip family we first use Eq.~\eqref{S:eq:matrixnecessary} to constrain the local contour, and then construct a real diagonal frame in which the full phase sum can be evaluated.

\subsection{Equivalent phase conditions from form-field convergence}

For completeness, we recall why Eq.~\eqref{S:eq:ksw} is the natural condition.  In a real diagonal frame, write
\begin{equation}
 g_{\alpha \beta}=\Lambda_\alpha\delta_{\alpha \beta},\qquad
 \sqrt{g}=\prod_{\alpha=1}^{D}\Lambda_\alpha^{1/2},
 \qquad -\pi<\Arg\Lambda_\alpha\leq\pi.\nn
\end{equation}
For a real scalar, the Gaussian kinetic term along the $\alpha$th direction is controlled by
\begin{equation}
 \sqrt{g}\,g^{\alpha\alpha}=\frac{\prod_\beta\Lambda_\beta^{1/2}}{\Lambda_\alpha}.\nn
\end{equation}
More generally, for a real $q$-form whose nonzero component lies along a set $I=\{\alpha_1,\ldots,\alpha_q\}$, the corresponding quadratic coefficient is proportional to
\begin{equation}
 \sqrt{g}\,g^{\alpha_1\alpha_1}\cdots g^{\alpha_q\alpha_q}.\nn
\end{equation}
Requiring a positive real part for all such coefficients is equivalent to demanding that the sum of the absolute eigenvalue phases be smaller than $\pi$. 
The closure $\Theta=\pi$ contains the Lorentzian real sections encountered below.  

Two qualifications will be used repeatedly.  First, the phases must be evaluated after a real change of basis on the chosen cycle.  A complex diagonalization can rotate the integration contour and is not a coordinate redundancy of the real functional integral.  Second, Eq.~\eqref{S:eq:matrixnecessary} is only a necessary test unless a real diagonal frame has been constructed.  This distinction is immaterial for the three-dimensional and hyperbolic examples, where the metrics are explicitly diagonal, but it is essential for the planar strip.

\section{One complex family metric for Poincar\'e $\AdS_3$, global $\AdS_3$, and BTZ}
\label{sec:ads3}

\subsection{Poincar\'e realization}

In Poincar\'e coordinates,
\begin{equation}
 \dd s^2=\frac{L^2}{z^2}\left(-\dd t^2+\dd x^2+\dd z^2\right),
 \label{S:eq:ads3poincare}
\end{equation}
a geodesic anchored at $(t,x)=(\pm t_0,\pm x_0)$ with $t_0,x_0>0$ can be written as
\begin{equation}
 z=\frac{\sqrt{x_0^2-t_0^2}}{\cosh\lambda},\qquad
 t=t_0\tanh\lambda,\qquad x=x_0\tanh\lambda.
 \label{S:eq:ads3embed}
\end{equation}
For spacelike separation the boundary is reached at $\lambda\to\pm\infty$. For timelike separation choose $\sqrt{x_0^2-t_0^2}=i\rho_t$, where $\rho_t=\sqrt{t_0^2-x_0^2}$, the boundary is reached at
\begin{equation}
 \lambda_L=-\infty-\frac{i\pi}{2},\qquad
 \lambda_R=+\infty+\frac{i\pi}{2},
\end{equation}

Differentiating Eq.~\eqref{S:eq:ads3embed} and using $\lambda=\lambda(u)$ gives
\begin{align}
 \frac{\dd s^2}{L^2}={}&\lambda'^2\dd u^2
 +\frac{t_0^2+(t_0^2-x_0^2)\sinh^2\lambda}
 {(t_0^2-x_0^2)^2}\,\dd t_0^2
 \nonumber\\
 &+\frac{x_0^2-(t_0^2-x_0^2)\sinh^2\lambda}
 {(t_0^2-x_0^2)^2}\,\dd x_0^2
 -\frac{2t_0x_0}{(t_0^2-x_0^2)^2}\,\dd t_0\dd x_0.
 \label{S:eq:ads3rawpullback}
\end{align}
The absence of $\dd u\,\dd t_0$ and $\dd u\,\dd x_0$ terms is a consequence of choosing $\lambda$ as an affine parameter along every geodesic.  

For the timelike family we define
\begin{equation}
 t_0=\rho_t\cosh\eta,\qquad x_0=\rho_t\sinh\eta,
 \qquad \xi=\log\rho_t, \nn
\end{equation}
and for the spacelike family we define
\bea
t_0=\rho_s\sinh\eta,\qquad x_0=\rho_s\cosh\eta,\qquad \xi=\log\rho_s.\nn
\eea
 Then replacing $\lambda$ by a contour $\lambda(u)$ gives
\begin{equation}
 \frac{ds^2_{\mathbb{C}}}{L^2_{\text{AdS}}}
 =\lambda'^2\dd u^2+\cosh^2\lambda\,\dd\xi^2
 -\sinh^2\lambda\,\dd\eta^2,
 \label{S:eq:ads3familymetric}
\end{equation}
where $L_{\text{AdS}}$ is the radius of AdS.
The spacelike and timelike cases differ only by their endpoint conditions in the complex $\lambda$ plane.

\subsection{BTZ realization}

For the nonrotating BTZ metric
\begin{equation}
 \frac{\dd s^2}{L^2_{\text{AdS}}}=-(r^2-r_h^2)\dd t^2
 +\frac{\dd r^2}{r^2-r_h^2}+r^2\dd\varphi^2,\nn
\end{equation}
where $0<r<+\infty$, $-\infty<t<+\infty$ and $0<\varphi<2\pi $, $r=r_h$ is the horizon.  We would like to consider the extremal line anchored at $(t,r,\varphi)=(t_0,+\infty,\pi+\varphi_0)$ and $(-t_0,+\infty,\pi-\varphi_0)$ with $t_0\in (0,+\infty)$ and $\varphi_0\in (0,\pi)$.
one convenient family of geodesics is
\begin{align}
 t&=\frac{1}{r_h}\text{arctanh}[-\tanh(r_ht_0)\tanh\lambda],\nonumber\\
 r&=r_h\sqrt{\frac{\cosh^2(r_h\varphi_0)+\sinh^2\lambda}
 {\cosh^2(r_h\varphi_0)-\cosh^2(r_ht_0)}},\nonumber\\
 \varphi&=\pi+\frac{1}{r_h}\text{arctanh}[-\tanh(r_h\varphi_0)\tanh\lambda].\nn
\end{align}
For the spacelike intervals $t_0<\varphi_0$, 
$\lambda \in (-\infty,+\infty)$, $\lambda\to \mp \infty$  corresponds to boundary 
condtions $(t,r,\varphi)=(\mp t_0,+\infty,\pi\mp \varphi_0)$. For the timeliek case $t_0>\varphi_0$, $\lambda\to \mp \infty \mp i\frac{\pi}{2}$ corresponds to the boundary conditions $(t,r,\varphi)=(\mp t_0,+\infty,\pi\mp \varphi_0)$.

Before making the final real transformation, the family pullback in $(u,t_0,\varphi_0)$ is
\begin{align}
 \frac{\dd s^2}{L^2_{\text{AdS}}}={}&\lambda'^2\dd u^2
 +\frac{r_h^2}{4}\left[
 \frac{1}{\sinh^2 r_h(t_0-\varphi_0)}
 +\frac{1}{\sinh^2 r_h(t_0+\varphi_0)}\right.
 \nonumber\\[-1mm]
 &\left.\hspace{22mm}
 +\frac{2\cosh 2\lambda}
 {\sinh r_h(t_0-\varphi_0)\sinh r_h(t_0+\varphi_0)}
 \right]\dd t_0^2
 \nonumber\\
 &-\frac{r_h^2}{2}\left[
 \frac{1}{\sinh^2 r_h(t_0-\varphi_0)}
 -\frac{1}{\sinh^2 r_h(t_0+\varphi_0)}
 \right]\dd t_0\dd\varphi_0
 \nonumber\\
 &+\frac{r_h^2}{4}\left[
 \frac{1}{\sinh^2 r_h(t_0-\varphi_0)}
 +\frac{1}{\sinh^2 r_h(t_0+\varphi_0)}\right.
 \nonumber\\[-1mm]
 &\left.\hspace{22mm}
 -\frac{2\cosh 2\lambda}
 {\sinh r_h(t_0-\varphi_0)\sinh r_h(t_0+\varphi_0)}
 \right]\dd\varphi_0^2.\nn
\end{align}
This formula makes clear that the dependence on the contour enters only through $\lambda'$ and $\cosh 2\lambda$.  The real transformation
\begin{equation}
 t_0=\frac{1}{r_h}\operatorname{arctanh}\frac{\cosh\xi}{\cosh\eta},
 \qquad
 \varphi_0=\frac{1}{r_h}\operatorname{arctanh}\frac{\sinh\xi}{\sinh\eta}
 \label{S:eq:btzfamilytransform}
\end{equation}
brings the family pullback exactly to Eq.~\eqref{S:eq:ads3familymetric}. Thus the horizon scale and thermal identification do not enter the local KSW phase problem.

\subsection{Global-$\AdS_3$ realization}

For
\begin{equation}
 \frac{\dd s^2}{L^2_{\text{AdS}}}=-(1+r^2)\dd t^2+\frac{\dd r^2}{1+r^2}
 +r^2\dd\varphi^2,
\end{equation}
with $r \in (0,+\infty)$, $t\in (-\infty,+\infty)$ and $\varphi\in (0,2\pi)$. The CFT lives on the boundary $r\to +\infty$. We would like to consider the geodeic lines  anchored at the boundary points $(t,r,\varphi)=(\pm t_0,+\infty,\pi\pm \varphi_0)$ with $t_0\in (0,+\infty)$ and $\varphi_0 \in (0,\pi)$.

A geodesic family may be parametrized as
\begin{align}
 r&=\sqrt{\frac{\cos^2\varphi_0+\sinh^2\lambda}
 {\sin^2\varphi_0-\sin^2 t_0}},\nonumber\\
 t&=-\arctan(\tan t_0\tanh\lambda),\nonumber\\
 \varphi&=\pi-\arctan(\tan\varphi_0\tanh\lambda).
 \label{S:eq:globalads3geodesic}
\end{align}
For the spacelike case $\varphi_0 >t_0$, $\lambda= \pm \infty$ corresponds to $(t,r,\varphi)=(\pm t_0,+\infty,  \pi\pm \varphi_0)$. While for the timelike case $t_0>\varphi_0$, $\lambda=\pm \infty \pm i\frac{\pi}{2}$ corresponds to the $(\pm t_0, +\infty, \pi \pm \phi_0)$.

The pullback before the final reparametrization is
\begin{align}
 \frac{\dd s^2}{L_{\text{AdS}}^2}={}&\lambda'^2\dd u^2
 +\frac14\left[
 \csc^2(t_0-\varphi_0)+\csc^2(t_0+\varphi_0)
 +2\cosh(2\lambda)\csc(t_0-\varphi_0)\csc(t_0+\varphi_0)
 \right]\dd t_0^2
 \nonumber\\
 &-\frac12\left[
 \csc^2(t_0-\varphi_0)-\csc^2(t_0+\varphi_0)
 \right]\dd t_0\dd\varphi_0
 \nonumber\\
 &+\frac14\left[
 \csc^2(t_0-\varphi_0)+\csc^2(t_0+\varphi_0)
 -2\cosh(2\lambda)\csc(t_0-\varphi_0)\csc(t_0+\varphi_0)
 \right]\dd\varphi_0^2.\nn
\end{align}
With the real endpoint transformation
\begin{equation}
 t_0=\frac{\pi}{2}+\arctan\frac{\sinh\xi}{\cosh\eta},
 \qquad
 \varphi_0=\arctan\frac{\sinh\eta}{\cosh\xi},
 \label{S:eq:globalfamilytransform}
\end{equation}
the pullback becomes Eq.~\eqref{S:eq:ads3familymetric}, up to the harmless interchange of the names of the two family coordinates.

Equations~\eqref{S:eq:ads3familymetric}, \eqref{S:eq:btzfamilytransform}, and \eqref{S:eq:globalfamilytransform} establish a stronger statement than formal similarity: Poincar\'e $\AdS_3$, global $\AdS_3$, and nonrotating BTZ induce the same complex symmetric bilinear form on their real family cycles. Since KSW is local and invariant under real changes of basis, all three backgrounds obey one contour theorem. The different global geometries only affect coordinate ranges, identifications, and which boundary anchors are represented.

\subsection{An inequality from KSW}

Write
\begin{equation}
\lambda(u)=g(u)+if(u),
\qquad g(u),f(u)\in\mathbb R,\nn
\end{equation}
and restrict attention to
\begin{equation}
-\frac{\pi}{2}\leq f(u)\leq\frac{\pi}{2},\nn
\end{equation}
as suggested by the endpoint conditions of the contour considered here. For the uniqueness argument, we assume that the contour is continuous and
piecewise $C^1$, with $g$ piecewise monotonic. Isolated vertical segments
are allowed, and the contour is restricted to the strip
$-\pi/2\leq f\leq\pi/2$. Self-retracing segments are regarded as
reparametrization redundancies and are therefore omitted. More general contours may leave this strip in the complex $\lambda$ plane,
potentially leading to different admissible cycles and hence to different
complex metrics. It would be interesting to investigate whether such
additional branches admit more physical interpretation.

The phase formulas can be checked directly from
\begin{align}
 \cosh(g+if)&=\cosh g\cos f+i\sinh g\sin f,\nonumber\\
 \sinh(g+if)&=\sinh g\cos f+i\cosh g\sin f.
 \label{S:eq:sinhdecomp}
\end{align}
For $|f|<\pi/2$, the real part of $\cosh\lambda$ is positive.  Hence
\begin{equation}
 |\Arg\cosh^2\lambda|
 =2\arctan|\tanh g\tan f|.\nn
\end{equation}
The extra minus sign in the third eigenvalue places $-\sinh^2\lambda$ close to the negative real axis when $f\to0$, and the principal absolute phase is
\begin{equation}
 |\Arg[-\sinh^2\lambda]|
 =\pi-2\arctan|\coth g\tan f|.\nn
\end{equation}
Finally, $\lambda'=g'+if'$ gives
\begin{equation}
 |\Arg\lambda'^2|=2\arctan|f'/g'|\nn
\end{equation}
on every nonvertical segment; vertical pieces are obtained by continuity.

Therefore, the absolute phases of the entries in Eq.~\eqref{S:eq:ads3familymetric} are
\begin{align}
 \phi_u&=2\arctan\left|\frac{f'}{g'}\right|,\nonumber\\
 \phi_\xi&=2\arctan|\tanh g\tan f|,\nonumber\\
 \phi_\eta&=\pi-2\arctan|\coth g\tan f|.
 \label{S:eq:ads3phases}
\end{align}
Therefore $\phi_u+\phi_\xi+\phi_\eta\leq\pi$ is equivalent to
\begin{equation}
 \left|\frac{f'}{g'}\right|
 \leq\left|\frac{\sin(2f)}{\sinh(2g)}\right|.
 \label{S:eq:diffineq}
\end{equation}

To see the last equivalence explicitly, set
\begin{equation}
 X=\left|\frac{f'}{g'}\right|,
 \qquad A=|\tanh g\tan f|,
 \qquad B=|\coth g\tan f|.\nn
\end{equation}
The phase inequality is
\begin{equation}
 \arctan X+\arctan A-\arctan B\leq0.\nn
\end{equation}
Since $B\geq A\geq0$ and the principal arctangent is increasing,
\begin{equation}
 \arctan B-\arctan A
 =\arctan\frac{B-A}{1+AB}.\nn
\end{equation}
A direct simplification gives
\begin{equation}
 \frac{B-A}{1+AB}
 =\left|\frac{\sin(2f)}{\sinh(2g)}\right|,\nn
\end{equation}
which proves Eq.~\eqref{S:eq:diffineq} without any loss of branch information in the AdS wedge.
On an interval where all relevant signs are fixed, define
\begin{equation}
 Q=\frac{|\tanh g|}{|\tan f|^\sigma},
 \qquad
 \sigma=\operatorname{sgn}(gg')\operatorname{sgn}(ff').
 \label{S:eq:Qdef}
\end{equation}
Then
\begin{equation}
 \frac{\dd}{\dd u}\log Q
 =2\operatorname{sgn}(gg')\left[
 \frac{|g'|}{|\sinh(2g)|}-\frac{|f'|}{|\sin(2f)|}\right].
 \label{S:eq:Qderivative}
\end{equation}
 Eq.~\eqref{S:eq:diffineq} implies
\begin{equation}
 \frac{|f'|}{|\sin2f|}\leq
 \frac{|g'|}{|\sinh2g|},
 \label{S:eq:ads3separatedineq}\nn
\end{equation}
which fixes the sign of the bracket and hence the monotonicity of $Q$ with a given sign of $gg'$.

\subsection{Common contour theorem}

  It is useful to formulate the endpoint argument as a first-departure lemma: if a nonnegative monotone quantity has limiting value zero at an asymptotic endpoint and its permitted monotonic direction points toward smaller values as one moves into the contour, it must vanish identically until a sign-changing point is reached.  Applying the lemma on successive sign-fixed intervals extends the conclusion to the entire half-contour.

Let us first consider the spacelike case. The endpoint conditions require
$f(u)\to0$ as $g(u)\to\pm\infty$. Suppose that the contour departs from
the real axis near the left endpoint. Then, on the first sign-definite
interval $g<0$ and $g'>0$, after the departure $f$ and $f'$ have the same sign, so that
$\sigma=-1$ and
\begin{equation}
Q(u)=|\tanh g(u)|\,|\tan f(u)|.\nn
\end{equation}
Since $f\to0$ as $g\to-\infty$, one has $Q\to0$ at the endpoint.
However, Eq.~\eqref{S:eq:Qderivative} implies that $Q$ is nonincreasing
along this branch. Because $Q$ is nonnegative, it must remain identically
zero, and hence $f=0$. The same argument excludes any later departure
from the real axis and therefore gives $f=0$ throughout the region
$g<0$. Repeating the argument from the right endpoint yields the same
conclusion for $g>0$. Thus the only admissible spacelike contour in the
class considered is the real contour $\lambda=g$, which reproduces the
standard Lorentzian AdS section.

For timelike case, the endpoint conditions are
\begin{equation}
 g\to-\infty,\quad f\to-\frac{\pi}{2}
 \qquad \text{and}\qquad
 g\to+\infty,\quad f\to+\frac{\pi}{2}.\nn
\end{equation}
The same first-departure argument gives $f=-\pi/2$ for $g<0$ and $f=+\pi/2$ for $g>0$. Continuity requires a segment at $g=0$. More explicitly, consider the left half $g<0$.  For spacelike data, suppose $f$ first becomes positive.  Immediately after the first departure, $f>0$, $f'>0$, and $Q=|\tanh g\tan f|$ tends to zero at $g=-\infty$.  Equation~\eqref{S:eq:Qderivative} requires $Q$ to decrease as $g$ increases toward the origin, which is impossible unless $Q=0$.  If the first departure is toward negative $f$, the same conclusion follows with $|f|$.  The right half is identical.  Hence the only regular spacelike branch is $f=0$.

For timelike data on the left, $f\to-\pi/2$.  A departure toward the interior of the strip has $f<0$ and $f'>0$.  The appropriate nonnegative quantity is $Q=|\tanh g|/|\tan f|$, which again tends to zero at the endpoint and is forced to decrease toward larger $g$.  It therefore remains zero, which means $|\tan f|=\infty$ and hence $f=-\pi/2$.  The right half similarly gives $f=+\pi/2$.  No finite-$g$ horizontal departure is possible, while continuity of the full path requires a connecting segment on the imaginary axis.  Up to orientation and regular reparametrization,
\begin{equation}
 \cC_{\AdS}:\qquad
 \lambda=\begin{cases}
 g-i\pi/2,&g<0,\\
 if,&g=0, \,f\in(-\pi/2,\pi/2),\\
 g+i\pi/2,&g>0.
 \end{cases}
 \label{S:eq:threepiece}
\end{equation}
On the horizontal branches $(\phi_u,\phi_\xi,\phi_\eta)=(0,\pi,0)$ and
\begin{equation}
 \frac{ds^2_{\text{hor}}}{L^2}
 =\dd g^2-\sinh^2g\,\dd\xi^2+\cosh^2g\,\dd\eta^2.
\label{S:eq:hor}
\end{equation}
On the middle part $(\phi_u,\phi_\xi,\phi_\eta)=(\pi,0,0)$ and
\begin{equation}
 \frac{\mathrm{d}s^2_{\text{mid}}}{L^2}
 =-\dd f^2+\cos^2f\,\dd\xi^2+\sin^2f\,\dd\eta^2.
\label{S:eq:mid}
\end{equation}
Thus the KSW-selected cycle is the same in Poincar\'e $\AdS_3$, global $\AdS_3$, and BTZ, and every open segment is a Lorentzian real section of one complex metric.

\subsection{Universal Lorentzian geometry in the three-dimensional AdS examples}
\label{sec:universal_ads3_segment}

A universal feature of the three-dimensional AdS examples considered in this work is that Poincar'e AdS$_3$, global AdS$_3$, and BTZ lead to the same local complex metric. This follows after appropriate real redefinitions of the parameters labeling the extremal-surface family:
\begin{equation}
\frac{\mathrm{d}s_{\mathbb{C}}^{2}}{L_{\text{AdS}}^{2}}
=
\mathrm{d}\lambda^{2}
+\cosh^{2}\lambda\,\mathrm{d}\xi^{2}
-\sinh^{2}\lambda\,\mathrm{d}\eta^{2},
\label{eq:universal_complex_ads3_metric}
\end{equation}
with $\lambda$ given by Eq.~\eqref{S:eq:threepiece}.
The differences among these examples are therefore encoded in the
coordinate ranges and global identifications, rather than in the local
complex geometry relevant to the KSW analysis.

On the two horizontal branches, the metric becomes (\ref{S:eq:hor}).
For the Poincar\'e AdS$_3$ construction, these branches should be
identified with the portions of the original Poincar\'e Lorentzian
section covered by the extremal-surface family. They extend from the two
boundary endpoints toward two components of the Poincar\'e horizon at
$z\rightarrow+\infty$.

Although Eq.~\eqref{S:eq:hor} can locally be rewritten
in the standard AdS-Rindler exterior form, this local coordinate
equivalence should not be interpreted as an identification of the
original Poincar\'e section with the complete AdS-Rindler exterior.
The Poincar\'e and AdS-Rindler patches are different global regions of
AdS$_3$, even though they are locally isometric.

The middle metric (\ref{S:eq:mid}) is locally pure AdS$_3$, rather than a new local
geometry. Consider the embedding of AdS$_3$ into $\mathbb{R}^{2,2}$,
with ambient metric
\begin{equation}
\mathrm{d}s_{\mathbb{R}^{2,2}}^{2}
=
-\mathrm{d}X_{-1}^{2}
-\mathrm{d}X_{0}^{2}
+\mathrm{d}X_{1}^{2}
+\mathrm{d}X_{2}^{2},
\end{equation}
and hyperboloid constraint
\begin{equation}
-X_{-1}^{2}
-X_{0}^{2}
+X_{1}^{2}
+X_{2}^{2}
=
-L_{\text{AdS}}^{2}.
\label{eq:ads3_embedding_constraint}
\end{equation}
The parametrization
\begin{align}
X_{-1}
&=
L\cos f\,\cosh\xi,
&
X_{1}
&=
L\cos f\,\sinh\xi,
\nonumber\\
X_{0}
&=
L\sin f\,\cosh\eta,
&
X_{2}
&=
L\sin f\,\sinh\eta\nn
\end{align}
satisfies Eq.~\eqref{eq:ads3_embedding_constraint}, and its induced
metric is precisely Eq.~\eqref{S:eq:mid}. Consequently,
\begin{equation}
R_{\mu\nu}
=
-\frac{2}{L_{\text{AdS}}^{2}}g_{\mu\nu},
\qquad
R
=
-\frac{6}{L_{\text{AdS}}^{2}},\nn
\end{equation}
confirming that the middle region is locally AdS$_3$.

The middle metric can be identified locally with the interior of an
AdS-Rindler wedge. Introducing the signed coordinate
$ r=\sin f$ with $-1<r<1$, one obtains
\begin{equation}
\frac{\mathrm{d}s_{\mathrm{mid}}^{2}}{L_{\text{AdS}}^{2}}
=
-(r^{2}-1)\,\mathrm{d}\xi^{2}
+
\frac{\mathrm{d}r^{2}}{r^{2}-1}
+
r^{2}\,\mathrm{d}\eta^{2}.
\label{eq:ads3_rindler_interior_metric}
\end{equation}
For $|r|<1$, the coordinate $r$ is timelike, and
Eq.~\eqref{eq:ads3_rindler_interior_metric} has the local form of the
AdS-Rindler interior. The two endpoints $r=1, r=-1$ correspond to the two horizon components at which the middle real
section is analytically connected to the original Poincar\'e sections.

The locus $r=0$, or equivalently $f=0$, is not a curvature singularity.
Near this point,
\begin{equation}
\frac{\mathrm{d}s_{\mathrm{mid}}^{2}}{L_{\text{AdS}}^{2}}
=
-\mathrm{d}f^{2}
+\mathrm{d}\xi^{2}
+f^{2}\,\mathrm{d}\eta^{2}
+O(f^{2})\,\mathrm{d}\xi^{2}
+O(f^{4})\,\mathrm{d}\eta^{2}.\nn
\end{equation}
The $(f,\eta)$ sector is locally the Milne form of two-dimensional
Minkowski spacetime. The degeneration of the $\eta$ direction at
$r=0$ is therefore a coordinate degeneration rather than a physical
singularity.

\subsection{Interpretation of the geometry in Poincar\'e and BTZ}
\paragraph{Poincar\'e case}

The three segments of the KSW-selected contour admit a simple geometric interpretation. For the timelike case, we define $\rho_t := \sqrt{t_0^2 - x_0^2}$ and choose $\sqrt{x_0^2 - t_0^2} = i\rho_t$.
Recall that the extremal line is parametrized by Eq.~\eqref{S:eq:ads3embed}.
On the lower horizontal branch, $\lambda = g - \frac{i\pi}{2}$ with $-\infty < g < 0$, the embedding coordinates become
\begin{equation}
z = -\frac{\rho_t}{\sinh g}, \qquad t = t_0 \coth g, \qquad x = x_0 \coth g.\nn
\end{equation}
Since $g < 0$, one has $z > 0$. As $g \to -\infty$, the extremal line approaches the first boundary endpoint: $(z, t, x) \longrightarrow (0, -t_0, -x_0)$.
As $g \to 0^-$, one finds $$z \longrightarrow +\infty, \qquad |t|, |x| \longrightarrow +\infty,$$ and the extremal line reaches the Poincaré horizon.

Similarly, on the upper horizontal branch, $\lambda = g + \frac{i\pi}{2}$ with $0 < g < +\infty$,
one obtains
\begin{equation}
z = \frac{\rho_t}{\sinh g}, \qquad t = t_0 \coth g, \qquad x = x_0 \coth g.
\end{equation}
This branch starts from the Poincaré horizon at $g \to 0^+$ and approaches the second boundary endpoint: $(z, t, x) \longrightarrow (0, t_0, x_0)$.

On the middle segment, $\lambda = if$ with $-\frac{\pi}{2} < f < \frac{\pi}{2}$,
the original Poincaré embedding coordinates become complex:
\begin{equation}
z = \frac{i\rho_t}{\cos f}, \qquad t = i t_0 \tan f, \qquad x = i x_0 \tan f.
\end{equation}
However, in the extremal-surface-adapted coordinates, the metric on this segment is (\ref{S:eq:mid}).
This is a real Lorentzian AdS--Rindler interior section. The contour coordinate $f$ becomes timelike, while the two limits
$f \to \pm\frac{\pi}{2}$
correspond to the two horizons reached from the horizontal branches at $g \to 0^{\pm}$, respectively.

The extremal curves follow the schematic trajectory:
\begin{align}
&\text{Poincar\'e boundary}
\longrightarrow
\text{Poincar\'e horizon}
\longrightarrow
\text{middle Lorentzian patch}\nn \\
&\longrightarrow
\text{Poincar\'e horizon}
\longrightarrow
\text{Poincar\'e boundary}.
\label{eq:ads3_complete_geometry}
\end{align}
The two horizontal branches belong to the original Poincar\'e
Lorentzian section. The middle branch, by contrast, is another real
Lorentzian section of the complexified AdS$_3$ geometry, locally
isometric to an AdS-Rindler interior. It is analytically attached to the
original Poincar\'e section at the two components of the Poincar\'e
horizon.

The resulting geometry should therefore not be identified with the
maximally extended AdS-Rindler spacetime. Rather, it consists of two
portions of the original Poincar\'e section connected through an
additional real Lorentzian section with the local geometry of an
AdS-Rindler interior. No physical junction condition, thin shell, or
additional matter source is introduced: the different real sections
are joined through analytic continuation inside a single complexified
AdS$_3$ geometry.

The same local complex metric and middle Lorentzian section also appear
in global AdS$_3$ and BTZ. In those cases, however, the interpretation of
the horizontal branches must be made in terms of the corresponding
original global or quotient Lorentzian sections. The BTZ horizon radius,
temperature, and angular quotient affect the global identifications but
do not alter the local KSW contour or the universal middle geometry.

Finally, for fixed values of the family coordinates $\xi$ and $\eta$,
the extremal curve becomes timelike on the middle branch. This provides
a geometric origin for the imaginary contribution to timelike
entanglement entropy. The interpretation resembles the original
proposal in which the imaginary part is associated with a timelike
extremal segment. In the present construction, however, the timelike
segment does not lie in the original Poincar\'e Lorentzian section.
Instead, it emerges when a single complex extremal surface passes
through another real Lorentzian section of the complexified geometry.\\

\paragraph{BTZ case} 
We now clarify the geometric meaning of the KSW-selected complex extremal line in the BTZ geometry. Consider a pair of timelike-separated boundary points characterized by 
$t_0 > \varphi_0 > 0$,
and introduce
\bea
a := r_h \phi_0, \qquad b := r_h t_0, \qquad b > a.\nn
\eea
The corresponding extremal line is parametrized by the complex affine coordinate $\lambda$, with the radial coordinate given by
\bea
\frac{r^2}{r_h^2} = \frac{\cosh^2 a + \sinh^2 \lambda}{\cosh^2 a - \cosh^2 b}.\nn
\eea
The KSW criterion selects the same three-piece contour Eq.~\eqref{S:eq:threepiece}.
Although the metric adapted to this contour takes the universal AdS-Rindler form, its embedding into the original BTZ real section has a nontrivial global interpretation.

On either horizontal branch, $\lambda = g \pm i\pi/2$, the radial coordinate becomes
\bea
\frac{r^2}{r_h^2} = \frac{\cosh^2 g - \cosh^2 a}{\cosh^2 b - \cosh^2 a}.\nn
\eea
Starting from one asymptotic endpoint, $|g| \to \infty$, the extremal line initially lies in the ordinary BTZ exterior. As $|g|$ decreases, it reaches
\bea
|g| = b \qquad \Longleftrightarrow \qquad r = r_h,\nn
\eea
and therefore crosses the BTZ event horizon. For
$a < |g| < b$, one has $0 < r^2 < r_h^2$,
so the line propagates through the usual black-hole interior. It subsequently reaches
\[
|g| = a \qquad \Longleftrightarrow \qquad r = 0.
\]
Here $r=0$ is the causal singularity associated with the BTZ quotient.

The complex extremal line does not terminate at $r=0$. Continuing the parameter $g$ beyond this point gives
\[
|g| < a \qquad \Longrightarrow \qquad r^2 < 0.
\]
Thus the trajectory leaves the standard real BTZ section and enters a different real Lorentzian section of the complexified BTZ geometry. In the extremal-surface-adapted coordinates, this new section remains locally real and Lorentzian, even though the original BTZ coordinates $(r, t, \phi)$ are generally complex. The line then reaches
$g = 0$, which is the horizon of the adapted AdS-Rindler patch.

It is essential to distinguish this horizon from the BTZ event horizon. The BTZ horizon is located at $r = r_h \Longleftrightarrow |g| = b$,
whereas the adapted AdS-Rindler horizon is located at $g=0$, which lies beyond the $r=0$ surface when viewed in the analytically continued BTZ coordinates. The latter is therefore not the global event horizon of the BTZ black hole, but a horizon associated with the foliation of the complexified geometry by the family of extremal lines.

On the middle segment, $\lambda = if$, the adapted metric becomes again real and Lorentzian. This segment connects the two adapted AdS-Rindler horizons at $f = \pm \pi/2$. After traversing this middle Lorentzian patch, the extremal line enters the second real section of the complexified BTZ geometry and follows the reverse sequence: it reaches $r=0$, re-enters the ordinary BTZ black-hole interior, crosses the BTZ event horizon, and finally arrives at the second asymptotic boundary endpoint.

The complete trajectory may therefore be summarized schematically as
\begin{equation}
\begin{aligned}
\text{BTZ boundary}
&\longrightarrow
\text{BTZ quotient singularity}
\longrightarrow
\text{another real section of complexified BTZ}
\nonumber \\
&\longrightarrow
\text{ middle Lorentzian patch}
\longrightarrow
\text{another real section of complexified BTZ}\nonumber \\
&\longrightarrow
\text{BTZ quotient singularity}
\longrightarrow
\text{BTZ boundary}.
\end{aligned}
\label{eq:btz_complete_geometry}
\end{equation}

This picture shows that the timelike extremal line cannot be understood as a curve entirely contained in the ordinary real BTZ spacetime. Instead, it passes through the physical BTZ exterior and interior, reaches the quotient singularity, and is then continued through the complexified geometry into additional real Lorentzian sections. The equality between the BTZ and Poincaré adapted metrics should therefore be understood as a local statement: the KSW-selected contour has the same local AdS-Rindler structure, while its global embedding into the BTZ quotient is substantially different.

\subsection{Geodesic length and timelike entanglement entropy}
\label{S:sec:ads3timelikelength}
An important consistency check is whether the constructed geometry reproduces the correct timelike entanglement entropy.
We now evaluate the geodesic length directly in the complex geometry
constructed from the Poincaré family. The extremal curves are specified by constant values of $\xi$ and $\eta$. For fixed values of the family coordinates $\xi$ and $\eta$, Eq.~\eqref{S:eq:ads3familymetric} reduces to
\begin{equation}
\dd s_{\gamma}^2 = L_{\text{AdS}}^2\lambda'(u)^2\dd u^2 .\nn
\end{equation}
The corresponding complex length is therefore
\begin{equation}
\mathcal{L}_{\gamma} = L_{\text{AdS}}\int_{\gamma}\dd u\,\sqrt{\lambda'(u)^2}.\nn
\end{equation}
We choose the branch of the square root continuously along the oriented
contour such that $\sqrt{\lambda'(u)^2}=\lambda'(u)$.
It follows that
\begin{equation}
\mathcal{L}_{\gamma} = L_{\text{AdS}}\int_{\mathcal{C}}\dd\lambda,\nn 
\end{equation}
where $\mathcal{C}$ denotes the contour.
At a finite radial cutoff $z=\epsilon$, the endpoints are given by
\begin{equation}
\lambda_L=-a_\epsilon-\frac{i\pi}{2},
\qquad
\lambda_R=a_\epsilon+\frac{i\pi}{2},
\qquad
a_\epsilon = \operatorname{arcsinh}\frac{\rho_t}{\epsilon}.
\end{equation}
The KSW-selected contour consists of three pieces,
\begin{equation}
\mathcal{C}= \mathcal{C}_{-}\cup\mathcal{C}_{0}\cup\mathcal{C}_{+},
\end{equation}
where
\begin{align}
\mathcal{C}_{-}:&\quad
\lambda=g-\frac{i\pi}{2},
\qquad
g:-a_\epsilon\longrightarrow 0,
\nn \\
\mathcal{C}_{0}:&\quad
\lambda=if,
\qquad
f:-\frac{\pi}{2}\longrightarrow\frac{\pi}{2},
\nn\\
\mathcal{C}_{+}:&\quad
\lambda=g+\frac{i\pi}{2},
\qquad
g:0\longrightarrow a_\epsilon .
\label{S:eq:ads3regulatedcontour}
\end{align}
The two horizontal segments give real contributions,
$\mathcal{L}_{-} = \mathcal{L}_{+} = L_{\text{AdS}}a_\epsilon$.
On the middle segment, 
we have $\mathcal{L}_{0}= i\pi L_{\text{AdS}}$.
The total regulated length is consequently
\begin{equation}
\mathcal{L}_{\gamma} = L_{\text{AdS}}\left(2a_\epsilon+i\pi\right) = L_{\text{AdS}}\left(
2\operatorname{arcsinh}\frac{\rho_t}{\epsilon}
+i\pi
\right).
\nn
\end{equation}
Equivalently, this follows directly from
$\mathcal{L}_{\gamma}=L_{\text{AdS}}(\lambda_R-\lambda_L)$.
Applying the holographic entropy formula gives
\begin{equation}
S_{\mathrm{T}} = \frac{\mathcal{L}{\gamma}}{4G_N} = \frac{L_{\text{AdS}}}{2G_N}
\operatorname{arcsinh}\frac{\rho_t}{\epsilon}
+
\frac{i\pi L_{\text{AdS}}}{4G_N}.\nn
\end{equation}
Using the Brown--Henneaux relation $c=\frac{3L_{\text{AdS}}}{2G_N}$,
we obtain
\begin{equation}
S_{\mathrm{T}} = \frac{c}{3}
\operatorname{arcsinh}\frac{\rho_t}{\epsilon}
+
\frac{i\pi c}{6}.\nn
\end{equation}
The invariant timelike separation between the two boundary endpoints is $$
\ell_t := \sqrt{(\Delta t)^2-(\Delta x)^2} = 2\rho_t.$$
In the asymptotic limit $\epsilon\to0$, the timelike entanglement entropy
therefore becomes
\begin{equation}
S_{\mathrm{T}} = \frac{c}{3}\log\frac{\ell_t}{\epsilon}
+
\frac{i\pi c}{6}
+
O\left(\frac{\epsilon^2}{\ell_t^2}\right),
\label{S:eq:ads3timelikeentropy}
\end{equation}
in agreement with the timelike entanglement entropy in the vacuum
$\text{CFT}_2$ \cite{Doi:2023zaf}.
Although the total length can be expressed solely in terms of the two
regulated endpoints, the KSW-selected contour provides a useful geometric
decomposition. Its real part comes from the two asymptotic spacelike
segments, while the universal imaginary contribution originates entirely
from the middle timelike segment. Reversing the contour orientation, or
equivalently choosing the opposite boundary $i\epsilon$ prescription,
gives the complex-conjugate result.

\section{One complex family metric for $\dS_3$}
\label{sec:ds3}

\subsection{Extremal family and adapted metric}

Take global $\dS_3$ coordinate
\begin{equation}
 \frac{\dd s^2}{L_{\text{dS}}^2}=-\dd T^2+\cosh^2T
 (\dd\theta^2+\sin^2\theta\,\dd\varphi^2).
 \label{S:eq:ds3global}
\end{equation}
The CFT lives on the asymptotic boundary $T\to +\infty$, which is a sphere $S^2$. We would like to consider the extremal line anchored at the boundary $(t,\theta,\varphi)=(+\infty,\frac{\pi}{2}-\theta_0,\pi-\varphi_0)$ and $(+\infty,\frac{\pi}{2}+\theta_0,\pi+\varphi_0)$ with $\theta_0\in(0,\frac{\pi}{2})$ and $\varphi_0\in(0,\pi)$.

With
\begin{equation}
 \Delta=\sin^2\theta_0+\cos^2\theta_0\sin^2\varphi_0,\nn
\end{equation}
the complex extremal lines are
\begin{align}
 T&=\operatorname{arcsinh}\!\left(\frac{i\cosh\lambda}{\sqrt\Delta}\right),\nonumber\\
 \theta&=\arccos\!\left(\frac{\sin\theta_0\sinh\lambda}
 {\sqrt{\cosh^2\lambda-\Delta}}\right),\nonumber\\
 \varphi&=\pi+\arctan[-\tan\varphi_0\tanh\lambda].
 \label{S:eq:ds3extremal}
\end{align}
The endpoint conditions are
\begin{equation}
 \lambda\to-\infty+\frac{i\pi}{2},\qquad
 \lambda\to+\infty-\frac{i\pi}{2}.
 \label{S:eq:ds3bc}
\end{equation}
Introduce real family coordinates $(\chi,\eta)$ by
\begin{equation}
 \theta_0=\arccos\sqrt{1-\frac{\sin^2\eta}{\cosh^2\chi}},
 \qquad
 \varphi_0=\arccos\!\left[-\frac{\sinh\chi}
 {\sqrt{\cosh^2\chi-\sin^2\eta}}\right].
 \label{S:eq:ds3familytransform}
\end{equation}
Substitution into Eq.~\eqref{S:eq:ds3global} gives the complex metrics
\begin{equation}
 \frac{ds^2_{\mathbb{C}}}{L^2_{\text{dS}}} =-\lambda'^2\dd u^2-\cosh^2\lambda\,\dd\chi^2
 -\sinh^2\lambda\,\dd\eta^2.
 \label{S:eq:ds3familymetric}
\end{equation}
This is a complex Einstein metric with positive cosmological constant wherever the family map is nondegenerate.

The coordinate transformation can be verified in two stages.  First substitute Eq.~\eqref{S:eq:ds3extremal} into the global metric while keeping $(\lambda,\theta_0,\varphi_0)$ independent.  The mixed terms involving $\dd\lambda$ cancel because $\lambda$ is affine along the extremal line.  The remaining two-dimensional metric on the space of endpoints is then diagonalized by Eq.~\eqref{S:eq:ds3familytransform}.  The signs in Eq.~\eqref{S:eq:ds3familymetric} are important: all three diagonal entries carry a minus sign before the complex phases are resolved.  Consequently the principal phase cone is not obtained from the AdS result by merely reversing a scalar differential inequality.

\subsection{The full phase cone}

Write $\lambda=g+if$ and orient the contour so that $g$ is nondecreasing. Motivated by the endpoint conditions in Eq.~\eqref{S:eq:ds3bc}, we restrict the contour to the strip $-\frac{\pi}{2}<f<\frac{\pi}{2}$.
As in the AdS analysis, we do not consider more complicated contours that leave this region.

 In the region
\begin{equation}
 g<0,\qquad 0<f<\frac{\pi}{2},
\label{dS_left_region}
\end{equation}
the three absolute phases are
\begin{align}
 \vartheta_u&=\pi-2\gamma,
 &\gamma&:=\arctan\left|\frac{f'}{g'}\right|,\nonumber\\
 \vartheta_\chi&=\pi-2\alpha,
 &\alpha&:=\arctan(|\tanh g|\tan f),\nonumber\\
 \vartheta_\eta&=\pi-2\beta,
 &\beta&:=\arctan(|\coth g|\tan f).
 \label{S:eq:ds3phases}
\end{align}
The formulas in Eq.~\eqref{S:eq:ds3phases} follow from the similar analysis as AdS example, but now every eigenvalue has an additional minus sign.  In the region (\ref{dS_left_region}), $\cosh^2\lambda$ and $\sinh^2\lambda$ approach the positive real axis before multiplication by $-1$, and therefore
\begin{equation}
 |\Arg[-\lambda'^2]|=\pi-2\gamma,
 \quad |\Arg[-\cosh^2\lambda]|=\pi-2\alpha,
 \quad |\Arg[-\sinh^2\lambda]|=\pi-2\beta.
\end{equation}
Adding these phases gives $3\pi-2(\alpha+\beta+\gamma)\leq\pi$.  Hence the KSW condition is exactly
\begin{equation}
 \alpha+\beta+\gamma\geq\pi.
 \label{S:eq:ds3phasecone}
\end{equation}
This is the correct ``reversed'' form of the AdS constraint in the physical dS wedge. A useful necessary consequence is the reversed ratio inequality
\begin{equation}
 \left|\frac{f'}{g'}\right|\geq
 \left|\frac{\sin2f}{\sinh2g}\right|.
 \label{S:eq:weakreverse}
\end{equation}
It is not equivalent to the full phase cone, and below it will be used
only after the branch information in Eq.~\eqref{S:eq:ds3phasecone} has
already supplied the relevant barrier and sign data.

\subsection{Barrier and monotonicity}
We first consider the left branch, which starts from $g\to-\infty, f\to\frac{\pi}{2}$.
Suppose that this branch departs from the horizontal line
$f=\pi/2$. Since the contour is restricted to
$-\pi/2\leq f\leq\pi/2$, the first departure must have $f'<0$.
The full phase condition restricts the region accessible to such a
departed branch. From
\begin{equation}
\tan\alpha\tan\beta=\tan^2f,\nn
\end{equation}
one finds
\begin{equation}
\alpha+\beta<\frac{\pi}{2},
\qquad
0<f<\frac{\pi}{4}.
\end{equation}
On a regular nonvertical segment, one also has $\gamma<\pi/2$.
Therefore, Eq.~\eqref{S:eq:ds3phasecone} cannot be satisfied for
$0<f<\pi/4$. At $f=\pi/4$, the KSW condition requires
$\gamma=\pi/2$, corresponding to a vertical tangent. Thus any departed
left branch must remain in the wedge
\begin{equation}
g<0,
\qquad
\frac{\pi}{4}\leq f<\frac{\pi}{2}.
\label{S:eq:ds3wedge}
\end{equation}
The branch also cannot reverse direction below $f=\pi/2$. Indeed, a
regular reversal would require $f'=0$ at finite $g$, and hence
$\gamma=0$. Equation~\eqref{S:eq:ds3phasecone} would then require
$\alpha+\beta\geq\pi$, which is impossible for finite $g$ and
$f<\pi/2$. Therefore, with the orientation $g'>0$, one has
$f'<0$ throughout the departed branch.
As in the AdS analysis, define on each interval with fixed signs
\begin{equation}
Q_{\dS}(u)
:=
\frac{|\tanh g(u)|}{|\tan f(u)|^{\sigma}},
\qquad
\sigma
:=
\operatorname{sgn}\bigl(g g'\bigr)
\operatorname{sgn}\bigl(f f'\bigr).
\label{S:eq:ds3Qgeneral}
\end{equation}
On the departed left branch,
\begin{equation}
g<0,
\qquad
g'>0,
\qquad
f>0,
\qquad
f'<0,
\end{equation}
so that $\sigma=1$. Hence
\begin{equation}
Q_{\dS}(u)
=
\frac{|\tanh g(u)|}{\tan f(u)}
=
|\tanh g(u)|\cot f(u).\nn
\end{equation}
On every regular nonvertical segment,
\begin{equation}
\frac{\dd}{\dd u}\log Q_{\dS}
=
2\operatorname{sgn}(gg')
\left[
\frac{|g'|}{|\sinh(2g)|}
-
\frac{|f'|}{|\sin(2f)|}
\right].\nn
\end{equation}
The reversed ratio inequality \eqref{S:eq:weakreverse} gives
\begin{equation}
\frac{|f'|}{|\sin(2f)|}
\geq
\frac{|g'|}{|\sinh(2g)|}.\nn
\end{equation}
Since $\operatorname{sgn}(gg')=-1$ on the left branch, it follows that
\begin{equation}
\frac{\dd}{\dd u}\log Q_{\dS}\geq0.\nn
\label{S:eq:ds3Q}
\end{equation}
Thus $Q_{\dS}$ is nondecreasing as the contour moves toward $g=0$.
The same conclusion holds on an isolated vertical segment: when
$g'=0$ and $f'<0$, Eq.~\eqref{S:eq:ds3Q} shows directly that
$Q_{\dS}$ increases as $f$ decreases.
At the left endpoint,
\begin{equation}
\lim_{g\to-\infty}Q_{\dS}=0,\nn
\end{equation}
because $f\to\pi/2$. On the other hand, the barrier
\eqref{S:eq:ds3wedge} implies $ 0\leq\cot f\leq1$.
Therefore, when a complete contour reaches the imaginary axis,
\begin{equation}
\lim_{g\to0^-}Q_{\dS}
=
\lim_{g\to0^-}|\tanh g|\cot f
=
0.\nn
\end{equation}
Since $Q_{\dS}$ is nonnegative and nondecreasing between two endpoints
where it vanishes, it must vanish identically. For $g<0$,
$|\tanh g|>0$, so this requires
\begin{equation}
f=\frac{\pi}{2},
\qquad
g<0.
\label{S:eq:ds3left}
\end{equation}
Hence the assumed departure from the horizontal branch is impossible.
\subsection{Right branch and the unique contour}
The family metric and the endpoint conditions are invariant under the
reflection
\begin{equation}
(g,f)\longmapsto(-g,-f).\nn
\end{equation}
Applying the preceding argument to the right branch gives
\begin{equation}
f=-\frac{\pi}{2},
\qquad
g>0.\nn
\end{equation}
Continuity then requires the two horizontal branches to be connected by
a segment on the imaginary axis, $g=0$. Thus, up to orientation and
regular real reparametrization, the unique contour is
\begin{equation}
\cC_{\dS}:\qquad
\lambda=
\begin{cases}
g+i\pi/2,
& g<0,
\\[2mm]
if, &g=0, f\in\left(-\frac{\pi}{2},\frac{\pi}{2}\right),
\\[2mm]
g-i\pi/2,
& g>0.
\end{cases}
\label{S:eq:ds3contour}
\end{equation}
On either horizontal branch, the family metric becomes
\begin{equation}
\frac{\dd s^2_{\mathrm{hor}}}{L_{\text{dS}}^2}
=
-\dd g^2
+\sinh^2g\,\dd\chi^2
+\cosh^2g\,\dd\eta^2,
\label{S:eq:ds3hor}
\end{equation}
for which
\begin{equation}
(\vartheta_u,\vartheta_\chi,\vartheta_\eta)
=
(\pi,0,0).\nn
\end{equation}
On the middle segment, $\lambda=if$, and the metric reduces to
\begin{equation}
\frac{\dd s^2_{\mathrm{mid}}}{L_{\text{dS}}^2}
=
\dd f^2
-\cos^2f\,\dd\chi^2
+\sin^2f\,\dd\eta^2,
\label{S:eq:ds3mid}
\end{equation}
with
\begin{equation}
(\vartheta_u,\vartheta_\chi,\vartheta_\eta)
=
(0,\pi,0).\nn
\end{equation}
Therefore, every nondegenerate segment saturates the KSW bound, and the
complete contour is assembled from Lorentzian real sections of the same
complexified $\dS_3$ geometry.
\subsection{Geometric interpretation of the complete dS$_3$ contour}
\label{sec:complete_ds3_geometry}

The KSW-selected contour decomposes the complexified dS$_3$ geometry
into three real Lorentzian segments. On the two horizontal branches,
$\lambda=g+i\pi/2$ with $g<0$ and $\lambda=g-i\pi/2$ with $g>0$, the
family metric reduces to (\ref{S:eq:ds3hor}).
Both branches belong to the same original future cosmological region of
dS$_3$, although they represent two different angular sectors in the
schematic diagram. They begin at two points on future infinity and
approach the cosmological horizons of an adapted static patch as
$g\rightarrow0$.

The relation to the global dS$_3$ coordinates can be made explicit by a
spatial rotation that places the relevant extremal curves on a fixed
$\phi$ plane, say $\phi=0$ or $\phi=\pi$. For the family of boundary
intervals with endpoints at
$\theta=\pi/2-\theta_0$ and $\theta=\pi/2+\theta_0$, the two outer
Lorentzian legs satisfy the condition that $(T,\theta)\longrightarrow(0,0)$ or $(T,\theta)\longrightarrow(0,\pi)$ when $g\to 0$,
independently of the value of $\theta_0$, for
$0<\theta_0\leq\pi/2$. These points are the two antipodal intersections
of the bifurcation circle with the chosen plane.

On the middle vertical branch, $\lambda=if$, the metric becomes
(\ref{S:eq:ds3mid}). Introducing the signed static radius $r=\sin f$
gives
\begin{equation}
\frac{\mathrm{d}s_{\mathrm{mid}}^{2}}{L_{\mathrm{dS}}^{2}}
=
-(1-r^{2})\,\mathrm{d}\chi^{2}
+\frac{\mathrm{d}r^{2}}{1-r^{2}}
+r^{2}\,\mathrm{d}\eta^{2},
\qquad
-1<r<1.
\label{eq:ds3_signed_static_patch}
\end{equation}
This region is locally the dS$_3$ static patch, extended smoothly from
one radial half to the other across the regular center $r=0$. Thus the
ranges $-1<r<0$ and $0<r<1$ should not be regarded as two independent
static patches: they are the two halves of one adapted static patch.
After suppressing the angular direction, they may be represented as two
triangular wedges joined at $r=0$, forming a square or diamond bounded
by the cosmological horizons at $r=\pm1$.

The adapted static Killing vector is $\partial_\chi$. Its norm is
\begin{equation}
|\partial_\chi|^2
=
\begin{cases}
L_{\mathrm{dS}}^2\sinh^2 g,
& \text{on the outer section},\\[1mm]
-L_{\mathrm{dS}}^2\cos^2 f,
& \text{on the middle section}.
\end{cases}
\end{equation}
It is spacelike on the outer section and timelike on the middle section,
while it becomes null at $\lambda=\pm i\frac{\pi}{2}$.
These loci correspond to the cosmological Killing horizons of the
adapted static observer. In a global-coordinate convention in which
this observer lies at $(\theta,\phi)=\left(\frac{\pi}{2},0\right)$,
the complete Killing horizons are
\begin{equation}
\tanh T=\pm\sin\theta\cos\phi .
\end{equation}
They are therefore the cosmological horizons of the static observer
located at the equatorial point $\theta=\pi/2$, rather than those of the
usual north-pole observer at $\theta=0$.

The coordinates $(g,\chi)$ and $(f,\chi)$ degenerate at the horizons,
but this is only a coordinate degeneration. By seeing the metric near the horizons, one could find the outer cosmological region
and the middle static region are smoothly connected across the full
null Killing horizons. In particular, other smooth curves can cross
the horizons at generic points. The extremal curves considered here
are more special. Since their family coordinates $\chi$ and $\eta$ are
held fixed and finite, they reach the bifurcation locus of the horizons,
rather than a generic point on a null generator.

For the dS$_3$ example, each extremal curve therefore follows the
schematic trajectory
\begin{align}
&\text{future infinity}
\longrightarrow
\text{cosmological horizon at its bifurcation locus}
\longrightarrow
\text{middle Lorentzian patch}
\nonumber\\
&\hspace{22mm}\longrightarrow
\text{cosmological horizon at its bifurcation locus}
\longrightarrow
\text{future infinity}.
\label{eq:ds3_complete_geometry}
\end{align}
The outer portions of the extremal curve are timelike, whereas the
middle portion is spacelike, even though the middle spacetime itself is
Lorentzian.

These regions should not be regarded as independent spacetimes joined
by a physical junction condition. There is no thin shell, localized
stress tensor, or additional junction source. Rather, the outer
cosmological section and the adapted static section are different real
Lorentzian sections of a single complexified dS$_3$ geometry. Their
intrinsic Lorentzian geometries admit a smooth extension across the full
null Killing horizons, while the KSW-selected extremal curves continue
analytically from one real section to the other through the
bifurcation locus.


\section{Hyperbolic extremal surfaces in $\AdS_4$}
\label{sec:hyperbolic}

\subsection{Derivation of the hyperbolic extremal surfaces in AdS$_4$}
\label{sec:derivation_hyperbolic_surface}

\begin{figure}[t]
\centering
\includegraphics[width=1\textwidth]{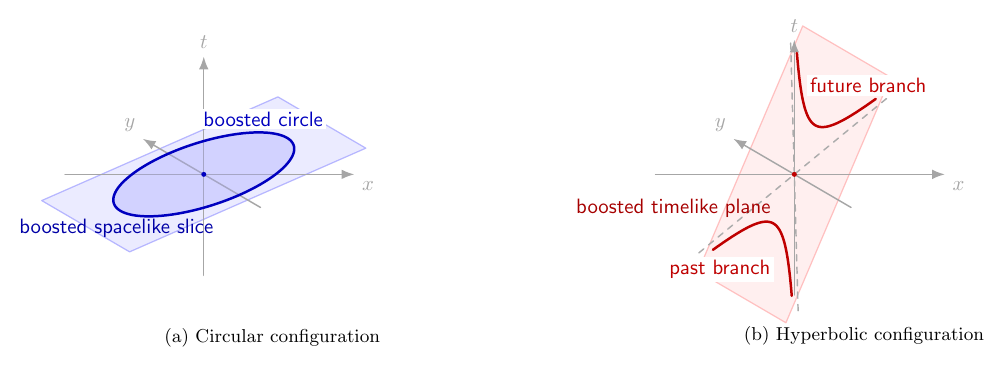}
\caption{
Boundary configurations in Poincar\'e AdS$_4$ obtained from the unified
quadratic construction.
(a) For $\Delta<0$, the entangling curve is a spatial circle on a
boosted Cauchy slice. The shaded disk denotes the corresponding boosted
ball.
(b) For $\Delta>0$, the boundary configuration consists of future and
past hyperbolic branches lying on a timelike plane. The dashed lines
denote the null directions in this plane.
The two configurations arise as different real sections of the same
complexified quadratic surface.
 }
 \label{fig:boosted-circle-hyperbola}
\end{figure}

In this subsection, we derive the codimension-two extremal surfaces associated with the spacelike circular and timelike hyperbolic boundary configurations.
We work in Poincar'e AdS$_4$,
\begin{equation}
ds^2
=
\frac{L_{\text{AdS}}^2}{z^2}
\left(
dz^2-dt^2+dx^2+dy^2
\right).
\label{eq:ads4_poincare_hyperbolic_derivation}
\end{equation}
Let $(t_0,x_0)$ be two real parameters and define
\begin{equation}
\Delta:=t_0^2-x_0^2.\nn
\end{equation}
We consider the two-dimensional plane
\begin{equation}
P_{t_0,x_0}:\qquad
t_0x-x_0t=0.
\label{eq:hyperbolic_plane}
\end{equation}
The boundary entangling curve is chosen to be
\begin{equation}
E_{t_0,x_0}:\qquad
t_0x-x_0t=0,
\qquad
t^2-x^2-y^2=\Delta.
\label{eq:unified_boundary_curve_derivation}
\end{equation}
The causal character of this configuration is determined by the sign of
$\Delta$. We shall derive the corresponding extremal surface separately
for the spacelike and timelike sectors. See Fig.~\ref{fig:boosted-circle-hyperbola} for an illustration.

\subsubsection{Spacelike circular subsystem}

We first consider
\begin{equation}
\Delta<0,
\qquad
\rho_{\mathrm{s}}
:=
\sqrt{x_0^2-t_0^2}.\nn
\end{equation}
It is convenient to parametrize $t_0=\rho_{\mathrm{s}}\sinh\beta, x_0=\rho_{\mathrm{s}}\cosh\beta$.
We then introduce the boosted coordinates
\begin{equation}
\widetilde t
:=
\frac{x_0t-t_0x}{\rho_{\mathrm{s}}},
\qquad
\widetilde x
:=
\frac{x_0x-t_0t}{\rho_{\mathrm{s}}}.
\label{eq:spacelike_boosted_coordinates}
\end{equation}
This is a real Lorentz transformation in the $(t,x)$ plane, and hence
\begin{equation}
-dt^2+dx^2
=
-d\widetilde t^{,2}
+d\widetilde x^{,2}.\nn
\end{equation}
The plane condition $P_{t_0,x_0}$ \eqref{eq:hyperbolic_plane} becomes $\widetilde t=0$,
while the boundary curve $E_{t_0,x_0}$ \eqref{eq:unified_boundary_curve_derivation}
reduces to
\begin{equation}
\widetilde t=0,
\qquad
\widetilde x^{,2}+y^2=\rho_{\mathrm{s}}^2.
\label{eq:spacelike_boundary_circle}
\end{equation}
Thus the spacelike boundary subsystem is an ordinary spatial circle in a
boosted Cauchy slice.

The rotational symmetry of the boundary circle motivates the general
bulk ansatz
\begin{equation}
\widetilde t=0,
\qquad
\widetilde x=R(z)\cos\theta,
\qquad
y=R(z)\sin\theta,\nn
\end{equation}
with boundary condition
\begin{equation}
R(0)=\rho_{\mathrm{s}}.\nn
\end{equation}
It is well-known the bulk extremal surface for the sphere is given by 
\begin{equation}
R(z)
=
\sqrt{\rho_{\mathrm{s}}^2-z^2},\nn
\end{equation}
or written as
\begin{equation}
\widetilde t=0,
\qquad
\widetilde x^{,2}+y^2+z^2
=
\rho_{\mathrm{s}}^2.
\label{eq:spacelike_hemisphere_standard}
\end{equation}
This is the usual hemispherical RT surface.

Returning to the original coordinates, the condition
$\widetilde t=0$ is equivalent to $t_0x-x_0t=0$.
Furthermore, on this plane,
\begin{equation}
t^2-x^2=-\widetilde x^{,2}.\nn
\end{equation}
Equation \eqref{eq:spacelike_hemisphere_standard} can therefore be written as
\begin{equation}
t^2-x^2-y^2-z^2
=
-\rho_{\mathrm{s}}^2
=
t_0^2-x_0^2.\nn
\end{equation}
Thus the spacelike extremal surface is
\begin{equation}
t_0x-x_0t=0,
\qquad
t^2-x^2-y^2-z^2=t_0^2-x_0^2.
\label{eq:spacelike_surface_boxed}
\end{equation}

Denote $t_0=\rho_{\mathrm{t}}\sinh\beta, x_0=\rho_{\mathrm{t}}\cosh\beta$, a convenient parametrization of the spacelike extremal surface is
\begin{align}
t&=t_0\cos\lambda\cos\theta,
\nonumber\\
x&=x_0\cos\lambda\cos\theta,
\nonumber\\
y&=\rho_{\mathrm{s}}\cos\lambda\sin\theta,
\nonumber\\
z&=\rho_{\mathrm{s}}\sin\lambda,
\label{eq:spacelike_surface_parametrization}
\end{align}
with
\begin{equation}
0\leq\lambda\leq\frac{\pi}{2},
\qquad
0\leq\theta<2\pi.
\label{eq:spacelike_parameter_range}
\end{equation}
At $\lambda=0$, the surface reaches the boundary circle, while
$\lambda=\pi/2$ corresponds to the tip of the hemisphere.

\subsubsection{Timelike hyperbolic subsystem}

We next consider
\begin{equation}
\Delta>0,
\qquad
\rho_{\mathrm{t}}
:=
\sqrt{t_0^2-x_0^2}.
\label{eq:rho_t_derivation}
\end{equation}
We parametrize $t_0=\rho_{\mathrm{t}}\cosh\beta, x_0=\rho_{\mathrm{t}}\sinh\beta$ and introduce the boosted coordinates
\begin{equation}
\tau
:=
\frac{t_0t-x_0x}{\rho_{\mathrm{t}}},
\qquad
\chi_{\perp}
:=
\frac{t_0x-x_0t}{\rho_{\mathrm{t}}}.\nn
\end{equation}
Again,
\begin{equation}
-dt^2+dx^2
=
-d\tau^2+d\chi_{\perp}^2.\nn
\end{equation}
The plane condition $P_{t_0,x_0}$ \eqref{eq:hyperbolic_plane} becomes
$\chi_{\perp}=0$, and the boundary curve $E_{t_0,x_0}$ \eqref{eq:unified_boundary_curve_derivation} becomes
\begin{equation}
\chi_{\perp}=0,
\qquad
\tau^2-y^2=\rho_{\mathrm{t}}^2.
\label{eq:timelike_boundary_hyperbola}
\end{equation}
It consists of a future and a past hyperbolic branch.

The future branch can be parametrized as
\begin{equation}
\tau=\rho_{\mathrm{t}}\cosh\eta,
\qquad
y=\rho_{\mathrm{t}}\sinh\eta,
\qquad
\eta\in\mathbb R.\nn
\end{equation}
The corresponding past branch is
\begin{equation}
\tau=-\rho_{\mathrm{t}}\cosh\eta,
\qquad
y=\rho_{\mathrm{t}}\sinh\eta.\nn
\end{equation}

The boundary configuration is invariant under boosts in the
$(\tau,y)$ plane. The most general bulk surface preserving this symmetry
may therefore be written as
\begin{equation}
\chi_{\perp}=0,
\qquad
\tau=R(z)\cosh\eta,
\qquad
y=R(z)\sinh\eta,\nn
\end{equation}
with
\begin{equation}
R(0)=\rho_{\mathrm{t}}.\nn
\end{equation}
Using $(z,\eta)$ as intrinsic coordinates, the induced metric becomes
\begin{equation}
ds_{\Sigma_{\mathrm{t}}}^2
=
\frac{L_{\text{AdS}}^2}{z^2}
\left[
\left(1-R'(z)^2\right)dz^2
+R(z)^2d\eta^2
\right].\nn
\end{equation}
The corresponding area functional is
\begin{equation}
\mathcal A_{\mathrm{t}}
=
L_{\text{AdS}}^2
\int d\eta\,dz \,
\frac{R(z)}{z^2}
\sqrt{1-R'(z)^2}.\nn
\end{equation}
The effective Lagrangian is
\begin{equation}
\mathcal L_{\mathrm{t}}
=
\frac{R}{z^2}\sqrt{1-R'^2},\nn
\end{equation}
and the Euler--Lagrange equation is
\begin{equation}
\frac{\sqrt{1-R'^2}}{z^2}
+
\frac{d}{dz}
\left[
\frac{RR'}{z^2\sqrt{1-R'^2}}
\right]
=0.\nn
\end{equation}

One could check the solution is given by
\begin{equation}
R(z)
=
\sqrt{\rho_{\mathrm{t}}^2+z^2}.
\end{equation}
 The extremal surface for the timelike case is thus
\begin{equation}
\chi_{\perp}=0,
\qquad
\tau^2-y^2-z^2=\rho_{\mathrm{t}}^2.\nn
\end{equation}

Returning to the original coordinates, $\chi_{\perp}=0$ gives $t_0x-x_0t=0$.
On this plane one has
\begin{equation}
t^2-x^2=\tau^2.\nn
\end{equation}
The extremal surface can therefore be written as
\begin{equation}
t_0x-x_0t=0,
\qquad
t^2-x^2-y^2-z^2=t_0^2-x_0^2.
\label{eq:timelike_surface_boxed}
\end{equation}

A convenient parametrization of the future real branch is
\begin{align}
t&=t_0\cosh\lambda\cosh\eta,
\nonumber\\
x&=x_0\cosh\lambda\cosh\eta,
\nonumber\\
y&=\rho_{\mathrm{t}}\cosh\lambda\sinh\eta,
\nonumber\\
z&=\rho_{\mathrm{t}}\sinh\lambda.
\label{eq:timelike_surface_parametrization}
\end{align}
For real $\lambda\geq0$, this parametrization reaches the future
boundary branch at $\lambda=0$. The past branch is obtained by analytic
continuation to
\begin{equation}
\lambda=i\pi,
\label{eq:timelike_past_endpoint}
\end{equation}
since
$\cosh(i\pi)=-1,\sinh(i\pi)=0$.
 Thus the algebraic extremal surface
connects the two boundary branches through the complexified
$\lambda$ plane. The extremal equation determines the complex quadratic
surface itself, whereas the choice of a particular real-dimensional contour
inside this complex surface requires an additional prescription, such as
the KSW criterion.

\subsubsection{Unified quadratic description}

The spacelike and timelike results take exactly the same algebraic form:
\begin{equation}
\Sigma_{t_0,x_0}:\qquad
t_0x-x_0t=0,
\qquad
t^2-x^2-y^2-z^2=t_0^2-x_0^2.
\label{eq:unified_extremal_surface_derived}
\end{equation}
The two sectors correspond to different real sections of this complexified
quadratic surface:
\begin{equation}
\begin{aligned}
t_0^2-x_0^2<0
&: \quad
\text{spacelike circular subsystem and real hemispherical RT surface},
\\
t_0^2-x_0^2>0
&: \quad
\text{timelike hyperbolic subsystem and complexified hyperbolic surface}.
\end{aligned}
\nn
\end{equation}

It remains to explain why solving the extremal equation inside the
three-dimensional planes used above is sufficient to establish extremality
in the full AdS$_4$ geometry. In the spacelike sector, the plane
$\widetilde t=0$ is the fixed-point set of the AdS isometry
\begin{equation}
\widetilde t\longrightarrow-\widetilde t.\nn
\end{equation}
Similarly, in the timelike sector, the plane $\chi_{\perp}=0$ is the
fixed-point set of
\begin{equation}
\chi_{\perp}\longrightarrow-\chi_{\perp}.\nn
\end{equation}
Both planes are therefore totally geodesic submanifolds of AdS$_4$.
The component of the mean-curvature vector normal to the corresponding
plane vanishes identically. The remaining component vanishes by the
Euler--Lagrange equations derived above. Hence the surfaces
\eqref{eq:spacelike_surface_boxed} and
\eqref{eq:timelike_surface_boxed} are genuine codimension-two extremal
surfaces in the full AdS$_4$ spacetime.

\subsection{Adapted metric and exact reduction}

Differentiating Eq.~\eqref{eq:timelike_surface_parametrization}, one finds the useful identities
\begin{align}
 -\dd t^2+\dd x^2+\dd y^2+\dd z^2
 =\rho_t^2\dd\lambda^2-\dd\rho_t^2+\rho_t^2\cosh^2\lambda
 \left(\dd\eta^2+\cosh^2\eta\,\dd\beta^2\right),\nn
\end{align}
while $z^2=\rho_t^2\sinh^2\lambda$.  Thus, using $(\lambda,\rho_t,\eta,\beta)$ as bulk coordinates, the metric is
\begin{equation}
 \frac{\dd s^2}{L_{\text{AdS}}^2}=
 \frac{\dd\lambda^2}{\sinh^2\lambda}
 -\frac{\dd\rho_t^2}{\rho_t^2\sinh^2\lambda}
 +\coth^2\lambda\left(\dd\eta^2+\cosh^2\eta\,\dd\beta^2\right).
 \label{S:eq:hypermetriclambda}
\end{equation}
Define
\begin{equation}
 \xi=\log\rho_t,
 \qquad
 \dd H_2^2=\dd\eta^2+\cosh^2\eta\,\dd\beta^2.\nn
\end{equation}
Then
\begin{equation}
 \frac{\dd s^2}{L_{\text{AdS}}^2}=
 \frac{\dd\lambda^2-\dd\xi^2}{\sinh^2\lambda}
 +\coth^2\lambda\,\dd H_2^2.\nn
\end{equation}
Introduce
\begin{equation}
 q=\log\tanh\frac{\lambda}{2}.
 \label{S:eq:qdef}
\end{equation}
The metric becomes
\begin{equation}
 \frac{\dd s^2}{L^2}=\dd q^2-\sinh^2q\,\dd\xi^2
 +\cosh^2q\,\dd H_2^2.\nn
\end{equation}
Finally, set
\begin{equation}
 \chi=q-\frac{i\pi}{2}.
 \label{S:eq:chidef}
\end{equation}
Since $\sinh(\chi+i\pi/2)=i\cosh\chi$ and $\cosh(\chi+i\pi/2)=i\sinh\chi$, one obtains
\begin{equation}
 \frac{\dd s^2}{L_{\text{AdS}}^2}=\dd\chi^2+\cosh^2\chi\,\dd\xi^2
 -\sinh^2\chi\,\dd H_2^2.
 \label{S:eq:hypermetricchi}
\end{equation}
For a contour $\chi=\chi(u)$ this is the $\AdS_3$ family metric with the single $\eta$ direction replaced by two identical hyperbolic directions.

\subsection{Endpoint map}
At a positive radial cutoff $z=\epsilon$, define
\begin{equation}
 \delta_\epsilon
 :=\operatorname{arcsinh}\!\left(\frac{\epsilon}{\rho_t}\right)>0,
 \qquad
 \lambda_L=\delta_\epsilon,
 \qquad
 \lambda_R=i\pi-\delta_\epsilon .\nn
\end{equation}
The right endpoint agrees with the chosen parametrization of the
past branch after the reparametrization $\eta\to-\eta$.
Let
\begin{equation}
 a_\epsilon=-\log\tanh\frac{\delta_\epsilon}{2}>0.\nn
\end{equation}
At the left endpoint,
\begin{equation}
 q_L=-a_\epsilon,
 \qquad
 \chi_L=-a_\epsilon-\frac{i\pi}{2}.\nn
\end{equation}
At the right endpoint, using the logarithm branch continuous along the desired surface,
\begin{equation}
 \tanh\frac{i\pi-\delta_\epsilon}{2}=-\coth\frac{\delta_\epsilon}{2},\nn
\end{equation}
so that
\begin{equation}
 q_R=a_\epsilon+i\pi,
 \qquad
 \chi_R=a_\epsilon+\frac{i\pi}{2}.\nn
\end{equation}
Thus $\epsilon\to0^+$ reproduces precisely the $\AdS_3$ timelike endpoint conditions.

\subsection{Full $\AdS_4$ KSW test}

Let
\begin{equation}
 \phi_u=|\Arg\chi'^2|,
 \qquad
 \phi_\xi=|\Arg\cosh^2\chi|,
 \qquad
 \phi_H=|\Arg(-\sinh^2\chi)|.
\end{equation}
Because $H_2$ has two directions with the same phase, the complete four-dimensional condition is
\begin{equation}
 \Theta_4=\phi_u+\phi_\xi+2\phi_H\leq\pi.
 \label{S:eq:theta4}
\end{equation}
It immediately implies
\begin{equation}
 \phi_u+\phi_\xi+\phi_H\leq\pi,
\end{equation}
which is exactly the $\AdS_3$ condition. Under the regularity assumptions used in the $\AdS_3$ contour theorem, necessity therefore fixes the image
\begin{equation}
 \chi=\begin{cases}
 g-i\pi/2,&g<0,\\
 if,&g=0,\quad f\in\left(-\frac{\pi}{2},\frac{\pi}{2} \right),\\
 g+i\pi/2,&g>0.
 \end{cases}
 \label{S:eq:hyperthreepiece}
\end{equation}
Sufficiency must still be checked because $\phi_H$ is counted twice. On the horizontal branches,
\begin{equation}
 \cosh^2(g\pm i\pi/2)=-\sinh^2g<0,
 \qquad
 -\sinh^2(g\pm i\pi/2)=\cosh^2g>0,\nn
\end{equation}
and $\chi'^2=g'^2>0$. Hence
\begin{equation}
 (\phi_u,\phi_\xi,\phi_H)=(0,\pi,0),
 \qquad \Theta_4=\pi.\nn
\end{equation}
On the vertical branch $\chi=if$,
\begin{equation}
 \chi'^2=-f'^2<0,
 \qquad
 \cosh^2(if)=\cos^2f>0,
 \qquad
 -\sinh^2(if)=\sin^2f>0,\nn
\end{equation}
and therefore
\begin{equation}
 (\phi_u,\phi_\xi,\phi_H)=(\pi,0,0),
 \qquad \Theta_4=\pi.\nn
\end{equation}
The full $\AdS_4$ KSW condition is thus saturated on every nondegenerate segment.

\subsection{Contour in the original $\lambda$ plane}

The inverse map follows from
\begin{equation}
 \chi=\log\tanh\frac{\lambda}{2}-\frac{i\pi}{2}.
 \label{S:eq:chilambda}
\end{equation}
The lower horizontal branch $\chi=g-i\pi/2$, $g<0$, gives $\lambda\in(0,+\infty)$. The middle branch $\chi=if$ gives
\begin{equation}
 \lambda=\log\cot\frac{\vartheta}{2}+\frac{i\pi}{2},
 \qquad \vartheta=f+\frac{\pi}{2}\in(0,\pi),\nn
\end{equation}
so $\Ima\lambda=\pi/2$ while $\Rea\lambda$ runs from $+\infty$ to $-\infty$. The upper horizontal branch maps to $\lambda=s+i\pi$ with $s<0$. Consequently
\begin{equation}
 \cC_\lambda:\quad
 \begin{cases}
 \lambda=s,&s:0\to+\infty,\\
 \lambda=s+i\pi/2,&s:+\infty\to-\infty,\\
 \lambda=s+i\pi,&s:-\infty\to0.
 \end{cases}
 \label{S:eq:originallambdacontour}
\end{equation}
The pieces join through the asymptotic regions $\Rea\lambda\to\pm\infty$. A direct vertical path $\lambda=i\vartheta$, $0<\vartheta<\pi$, is inequivalent: it gives two negative eigenvalues, $\Theta_4=2\pi$, and is excluded.

\subsection{Spacelike circular subsystem}

For $\Delta<0$, set
\begin{equation}
 \rho_s=\sqrt{x_0^2-t_0^2},
 \qquad
 t_0=\rho_s\sinh\beta,
 \qquad x_0=\rho_s\cosh\beta.\nn
\end{equation}
The bulk surface is the boosted hemisphere. A convenient parametrization leads to
\begin{equation}
 \frac{\dd s^2}{L_{\text{AdS}}^2}=
 \frac{\dd\lambda^2}{\sin^2\lambda}
 +\frac{\dd\rho_s^2}{\rho_s^2\sin^2\lambda}
 +\cot^2\lambda\left(\dd\theta^2-\cos^2\theta\,\dd\beta^2\right).
 \label{S:eq:spacelikeballmetric}
\end{equation}
For a complex contour $\lambda(u)$, the four eigenvalues are proportional to
\begin{equation}
 \frac{\lambda'^2}{\sin^2\lambda},\qquad
 \frac{1}{\sin^2\lambda},\qquad
 \cot^2\lambda,\qquad
 -\cot^2\lambda.
\label{S:eq:balleigenvalues}
\end{equation}
For any nonzero complex $Q$ on the principal branch,
\begin{equation}
 |\Arg Q|+|\Arg(-Q)|=\pi.\nn
\end{equation}
The last two entries in Eq.~\eqref{S:eq:balleigenvalues} already saturate KSW. The first two must therefore have vanishing phase. Regularity and connection to a positive real cutoff select
\begin{equation}
 0<\lambda\leq\frac{\pi}{2},\qquad \lambda\in\mathbb R,\nn
\end{equation}
which is precisely the ordinary RT hemisphere.

\subsection{Null degeneration}

At $t_0^2=x_0^2$, both $\rho_t$ and $\rho_s$ vanish and the adapted coordinates used in the two non-null sectors become singular.  The quadratic equation \eqref{eq:unified_extremal_surface_derived} remains meaningful, but its boundary curve degenerates onto null generators.  The null configuration can be approached from either side, yet the foliation by fixed-size hyperboloids or circles collapses.  Since the KSW test is a test of the pullback to a nondegenerate real cycle, the exactly null case requires a separate parametrization and is not included in the uniqueness statements above.

\section{Planar strips in general dimension}
\label{sec:stripgeneral}

\begin{figure}[t]
\centering
\includegraphics[width=1\textwidth]{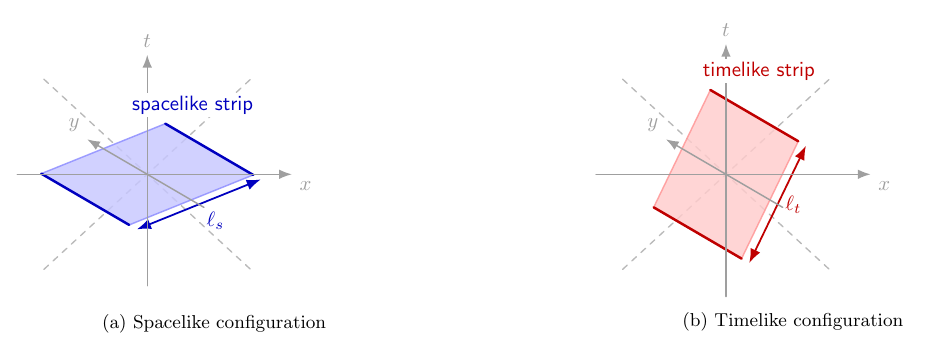}
\caption{
Tilted strip configurations in the boundary Minkowski spacetime.
(a) A spacelike strip with invariant width
$\ell_s=2\rho_s=2\sqrt{x_0^2-t_0^2}$.
(b) A timelike strip with invariant temporal width
$\ell_t=2\rho_t=2\sqrt{t_0^2-x_0^2}$.
Both strips are translationally invariant along the transverse $y$
direction, and the dashed lines denote the null directions in the
$(t,x)$ plane.
 }
 \label{fig:strip}
\end{figure}

\subsection{Unified complexified extremal-surface equation}

In Poincaré $\AdS_{d+1}$,
\begin{equation}
\dd s^2=\frac{L_{\text{AdS}}^2}{z^2}
\left(
-\dd t^2+\dd x^2+\dd z^2
+\dd\bm y_{d-2}^2
\right),
\label{S:eq:adsdmetric}
\end{equation}
consider two parallel boundary planes located at $(t,x)=(\pm t_0,\pm x_0)$,
with translational invariance in the transverse directions $\bm y$. See Fig.~\ref{fig:strip} for an illustration. We can treat the spacelike and timelike strip configurations within a unified framework. 

A codimension-two surface may be parametrized by
$(t(\sigma),x(\sigma),z(\sigma),\bm y)$, with area
\begin{equation}
\mathcal{A}
=
L_{\text{AdS}}^{d-1}V_\perp
\int\dd\sigma
\frac{
\sqrt{-\dot{t}^2+\dot{x}^2+\dot{z}^2}
}{z^{d-1}}.\nn
\end{equation}
The conserved momenta associated with translations in $t$ and $x$
imply
\begin{equation}
\frac{\dot{t}}{\dot{x}}
=
\frac{t_0}{x_0}.\nn
\end{equation}
Hence every surface lies in a fixed two-plane in the $(t,x)$ directions.
To treat spacelike and timelike endpoint separations uniformly, define 
$\Delta:=x_0^2-t_0^2, R:=\sqrt{\Delta}$,
where the square-root branch is chosen as
\begin{equation}
R=
\begin{cases}
\rho_s:=\sqrt{x_0^2-t_0^2},
& x_0^2>t_0^2, \\[1mm]
i\rho_t:=i\sqrt{t_0^2-x_0^2},
& t_0^2>x_0^2.
\end{cases}
\nn
\end{equation}
We first introduce an arbitrary normalization constant $C$ and write
\begin{equation}
t=\frac{t_0}{C}\lambda,
\qquad
x=\frac{x_0}{C}\lambda,
\qquad
z=\frac{R}{C}\zeta.
\label{S:eq:stripembeddingC}
\end{equation}
Using $\zeta$ as a surface parameter gives
\begin{equation}
-\left(\frac{\dd t}{\dd\zeta}\right)^2
+\left(\frac{\dd x}{\dd\zeta}\right)^2
+\left(\frac{\dd z}{\dd\zeta}\right)^2
=
\frac{\Delta}{C^2}
\left[
1+\left(\frac{\dd\lambda}{\dd\zeta}\right)^2
\right].
\label{S:eq:stripkinematicidentity}
\end{equation}
The area functional therefore reduces to
\begin{equation}
\mathcal{A}
=
L_{\text{AdS}}^{d-1}V_\perp
\left(\frac{C}{R}\right)^{d-2}
\int\dd\zeta
\frac{
\sqrt{1+(\partial_\zeta\lambda)^2}
}{\zeta^{d-1}}.
\label{S:eq:stripreducedarea}
\end{equation}
For timelike separation the prefactor in
Eq.~\eqref{S:eq:stripreducedarea} is generally complex. It is,
however, independent of $\lambda(\zeta)$ and therefore does not modify
the extremal-surface equation. This explains why the spacelike and
timelike sectors obey the same complexified differential equation.
Translation invariance in $\lambda$ gives, on the branch leaving the
endpoint $\lambda=C$,
\begin{equation}
\frac{\dd\lambda}{\dd\zeta}
=
-\frac{\zeta^{d-1}}
{\sqrt{1-\zeta^{2(d-1)}}},
\label{S:eq:firstintegral}
\end{equation}
and hence
\begin{equation}
C-\lambda=F_d(\zeta),
\qquad
F_d(\zeta):=
\int_0^\zeta
\frac{s^{d-1}\dd s}
{\sqrt{1-s^{2(d-1)}}}.
\label{S:eq:Fddef}
\end{equation}
The branch leaving the other endpoint is correspondingly described by
\begin{equation}
C+\lambda=F_d(\zeta).
\label{S:eq:secondstripbranch}
\end{equation}
For the ordinary spacelike strip, the two real branches meet smoothly
at
\begin{equation}
\lambda=0,
\qquad
\zeta=1.
\end{equation}
This fixes
\begin{equation}
C=c_d:=F_d(1)
=
\frac{1}{2(d-1)}
\mathrm{B}\left(
\frac{d}{2(d-1)},\frac{1}{2}
\right)
=
\frac{
\sqrt{\pi}\,\Gamma\left(\frac{d}{2(d-1)}\right)
}{
2(d-1)
\Gamma\left(\frac{2d-1}{2(d-1)}\right)
}.
\label{S:eq:cd}
\end{equation}
In what follows, we study the timelike family obtained by analytic
continuation of this standard connected strip and therefore retain the
same normalization constant $c_d$. The continuation changes the phase
of $R$ and the integration cycle on the Riemann surface of $F_d$, rather
than the dimensionless constant $c_d$ itself.
The unified embedding is thus
\begin{equation}
t=\frac{t_0}{c_d}\lambda,
\qquad
x=\frac{x_0}{c_d}\lambda,
\qquad
z=\frac{\sqrt{x_0^2-t_0^2}}{c_d}\zeta.
\label{S:eq:stripembedding}
\end{equation}
For spacelike separation the square root and the standard contour are
real. For timelike separation,
we choose $\sqrt{x_0^2-t_0^2}=i\rho_t$,
so that
\begin{equation}
t=\frac{t_0}{c_d}\lambda,
\qquad
x=\frac{x_0}{c_d}\lambda,
\qquad
z=\frac{i\rho_t}{c_d}\zeta,
\label{S:eq:timelikestripembedding}
\end{equation}
where both $\lambda$ and $\zeta$ are generally complex. The two
boundary endpoints correspond to
\begin{equation}
(\lambda,\zeta)=(c_d,0),
\qquad
(\lambda,\zeta)=(-c_d,0).
\end{equation}
Thus the spacelike and timelike surfaces solve the same complexified
extremal equation, but correspond to different integration cycles on
its Riemann surface.
\subsection{Spacelike and timelike family metrics}\label{section_metric_strip}
Let $u$ be a real parameter along the chosen contour, with
$\lambda=\lambda(u)$ and $\zeta=\zeta(u)$. For spacelike endpoint data,
we introduce
\begin{equation}
t_0=\rho_s\sinh\eta,
\qquad
x_0=\rho_s\cosh\eta,
\qquad
\rho_s^2=x_0^2-t_0^2.\nn
\end{equation}
For timelike endpoint data, we instead use
\begin{equation}
t_0=\rho_t\cosh\eta,
\qquad
x_0=\rho_t\sinh\eta,
\qquad
\rho_t^2=t_0^2-x_0^2.\nn
\end{equation}
Define
\begin{equation}
\epsilon=
\begin{cases}
+1, & \text{spacelike family}, \\
-1, & \text{timelike family},
\end{cases}
\qquad
\rho=
\begin{cases}
\rho_s, & \text{spacelike family}, \\
\rho_t, & \text{timelike family}.
\end{cases}
\end{equation}
A direct pullback of the Poincaré metric gives the unified family
metric
\begin{align}
\frac{\dd s^2}{L_{\text{AdS}}^2}
={}&
\frac{\lambda^2+\zeta^2}
{\rho^2\zeta^2}\dd\rho^2
+
\frac{2(\lambda\lambda'+\zeta\zeta')}
{\rho\zeta^2}\dd\rho\,\dd u
+
\frac{\lambda'^2+\zeta'^2}
{\zeta^2}\dd u^2
\nonumber \\
&-
\frac{\lambda^2}{\zeta^2}\dd\eta^2
+
\epsilon
\frac{c_d^2}{\rho^2\zeta^2}
\dd\bm y_{d-2}^2.
\label{S:eq:generalstripfamilymetric}
\end{align}
The $(u,\rho)$ block is therefore identical in the two causal sectors.
The difference is the sign of the transverse metric
$\dd\bm y_{d-2}^2$, which originates from the factor
$z^2=-\rho_t^2\zeta^2/c_d^2$ on the timelike branch.
In particular, the timelike family metric is
\begin{align}
\frac{\dd s_t^2}{L_{\text{AdS}}^2}
={}&
\frac{\lambda^2+\zeta^2}
{\rho_t^2\zeta^2}\dd\rho_t^2
+
\frac{2(\lambda\lambda'+\zeta\zeta')}
{\rho_t\zeta^2}\dd\rho_t\,\dd u
+
\frac{\lambda'^2+\zeta'^2}
{\zeta^2}\dd u^2
\nonumber \\
&-
\frac{\lambda^2}{\zeta^2}\dd\eta^2
-
\frac{c_d^2}{\rho_t^2\zeta^2}
\dd\bm y_{d-2}^2,
\label{S:eq:general-timelike-strip-metric}
\end{align}
whereas the spacelike family metric is
\begin{align}
\frac{\dd s_s^2}{L_{\text{AdS}}^2}
={}&
\frac{\lambda^2+\zeta^2}
{\rho_s^2\zeta^2}\dd\rho_s^2
+
\frac{2(\lambda\lambda'+\zeta\zeta')}
{\rho_s\zeta^2}\dd\rho_s\,\dd u
+
\frac{\lambda'^2+\zeta'^2}
{\zeta^2}\dd u^2
\nonumber \\
&-
\frac{\lambda^2}{\zeta^2}\dd\eta^2
+
\frac{c_d^2}{\rho_s^2\zeta^2}
\dd\bm y_{d-2}^2.
\label{S:eq:general-spacelike-strip-metric}
\end{align}
Although the differential equation relating $\lambda$ and $\zeta$ is
the same, the two induced family metrics are therefore different real
sections of the same complexified construction.

\subsection{Boundary phase from a positive real cutoff and the leading KSW condition}
\label{S:sec:boundaryphase}
We now determine the leading phase of the extremal-surface contour near the asymptotic boundary. We first impose the standard requirement that the holographic coordinate leaves the boundary along the positive real direction. This ensures that the dual CFT is regulated by a real positive ultraviolet cutoff. We then show that, for $d>2$, the same phase is independently selected by the leading-order KSW condition, up to the orientation of the radial direction.

For timelike separation, we have (\ref{S:eq:timelikestripembedding}).
Requiring $z$ to leave the conformal boundary along the positive real direction fixes
\begin{equation}
\zeta(u)=-iu+O(u^2),
\qquad
u\to0^+.
\label{S:eq:realcutoffzeta}
\end{equation}
Using the universal near-boundary extremal-surface relation
\begin{equation}
c_d-\lambda = F_d(\zeta) =\frac{\zeta^d}{d} + O\left(\zeta^{3d-2}\right),
\label{S:eq:leadingFdrelation}
\end{equation}
we obtain
\begin{equation}
\lambda(u) = c_d-\frac{(-i)^d}{d} u^d + O(u^{d+1}).
\label{S:eq:generalboundaryphase}
\end{equation}
Thus, the leading phase follows directly from the extremal-surface equation together with the positive real radial cutoff. It is not a freely chosen complex deformation.

More generally, consider the local expansions
\begin{equation}
\lambda(u) = c_d-\frac{e^{i\theta}}{d} u^d + \sum_{n\geq1}a_nu^{d+n},
\qquad
\zeta(u) = \omega u+b_2u^2+b_3u^3+\cdots,
\label{S:eq:generalboundaryexpansions}
\end{equation}
where $\theta\in\mathbb{R}$, while the coefficients $a_n$ are generally complex. The leading term in Eq.~\eqref{S:eq:leadingFdrelation} gives
\begin{equation}
\omega^d=e^{i\theta}.
\label{S:eq:omegatheta}
\end{equation}
The positive-cutoff condition~\eqref{S:eq:realcutoffzeta} selects
$\omega=-i$, and therefore $e^{i\theta}=(-i)^d$.
The first inverse coefficients are
\begin{equation}
b_2=-\omega^{1-d}a_1,
\qquad
b_3=-\omega^{1-d}a_2 - \frac{d-1}{2} \omega^{1-2d}a_1^2.
\label{S:eq:inversecoefficients}
\end{equation}
We next show that the leading KSW condition provides an independent constraint on the same phase. Introduce a positive real radial coordinate
\begin{equation}
r:=\frac{\rho_t}{c_d} u,
\qquad
r>0.\nn
\end{equation}
Near the boundary, Eq.~\eqref{S:eq:generalboundaryexpansions} implies
\begin{equation}
z=i\omega r+O(r^2),
\qquad
t=\rho_t\cosh\eta+O(r^d),
\qquad
x=\rho_t\sinh\eta+O(r^d).
\end{equation}
Consequently, the leading bulk metric takes the diagonal form
\begin{equation}
ds^2 = \frac{L_{\text{AdS}}^2}{r^2} \left[ dr^2 + (i\omega)^{-2} \left( -d\rho_t^2 + \rho_t^2d\eta^2 + d\vec{y}_{d-2}^2 \right) \right] + O(r^{-1}).
\label{S:eq:leadingmetricKSW}
\end{equation}
The subleading terms do not affect the leading phase analysis.

Up to positive real factors, the eigenvalues of the leading metric are
\begin{equation}
1,
\qquad
-(i\omega)^{-2},
\qquad
(i\omega)^{-2},
\end{equation}
where the last eigenvalue is repeated $(d-1)$ times. Define
\begin{equation}
\varphi := \operatorname{Arg}\left[(i\omega)^{-2}\right],
\qquad
-\pi\leq\varphi\leq\pi.
\end{equation}
Since $\left| \operatorname{Arg}\left[-e^{i\varphi}\right] \right| = \pi-|\varphi|$,
the leading KSW phase sum is
\begin{align}
\Theta_{\mathrm{KSW}}^{(0)} &= \left| \operatorname{Arg}\left[-(i\omega)^{-2}\right] \right| + (d-1) \left| \operatorname{Arg}\left[(i\omega)^{-2}\right] \right| \nonumber \\
&= \pi-|\varphi|+(d-1)|\varphi| \nonumber \\
&=  \pi+(d-2)|\varphi| .
\label{S:eq:leadingKSWsum}
\end{align}
Including the Lorentzian limiting case, the KSW condition requires
\begin{equation}
\Theta_{\mathrm{KSW}}^{(0)}\leq\pi.
\end{equation}
For $d>2$, Eq.~\eqref{S:eq:leadingKSWsum} therefore forces $\varphi=0$.
Since Eq.~\eqref{S:eq:omegatheta} implies $|\omega|=1$, it follows that
\begin{equation}
i\omega=\pm1,
\qquad\text{or equivalently}\qquad
\omega=\mp i.
\end{equation}
Thus, the leading KSW condition forces the holographic coordinate to leave the boundary along a real radial direction:
\begin{equation}
z=\pm r+O(r^2).
\end{equation}
The standard orientation of the Poincaré radial coordinate, namely $z>0$ for $r>0$, selects $i\omega=1,\omega=-i$.
Combining this result with Eq.~\eqref{S:eq:omegatheta}, we again obtain
\begin{equation}
e^{i\theta}=(-i)^d.
\label{S:eq:thetafromKSW}
\end{equation}
The positive real cutoff argument and the leading KSW analysis therefore provide two complementary ways of determining the boundary phase. The cutoff condition directly selects the positive real radial branch, whereas the KSW condition first restricts the contour to the two real radial orientations ($z\simeq\pm r$). The physical choice ($z>0$) then reproduces the same phase.

There is a minor distinction between even and odd $d$. For even $d$, the two KSW-allowed choices ($\omega=\pm i$) give the same value of $\omega^d$. Hence, the leading KSW condition alone already uniquely fixes $e^{i\theta}=(-i)^d$ for $d$ being even.
For odd $d$, the two choices differ by the orientation of the radial coordinate, and the positive-cutoff condition is required to select the physical sign.
The case $d=2$ is exceptional. In this case, $\Theta_{\mathrm{KSW}}^{(0)}=\pi$,
for arbitrary $\varphi$, so the leading local KSW condition does not constrain the phase $\omega$. The positive real cutoff still fixes $\omega=-i$, but a determination based only on the KSW criterion requires either subleading information or the full global contour analysis. This degeneracy is consistent with the special role of the global KSW monotonicity argument in $\text{AdS}_3$.

In summary, for a timelike strip one finds
\begin{equation}
\zeta(u)=-iu+O(u^2),
\qquad
e^{i\theta}=(-i)^d,
\qquad
\lambda(u) = c_d-\frac{(-i)^d}{d} u^d + O(u^{d+1}).
\label{S:eq:boundaryphasesummary}
\end{equation}
For $d>2$, this result is supported both by the positive real cutoff condition and by leading-order KSW admissibility.

\subsection{Even--odd structure of the straight timelike ray}

The requirement of a positive real radial cutoff selects the local behavior
\begin{equation}
\zeta=-iu+O(u^2)
\end{equation}
near the conformal boundary. It is therefore natural to examine the exact
straight continuation
\begin{equation}
\zeta=-iu,
\qquad
u>0.
\label{S:eq:straighttimelikeray}
\end{equation}
Along this ray, Eq.~\eqref{S:eq:Fddef} becomes
\begin{equation}
F_d(-iu)
=
(-i)^d
\int_0^u
\frac{v^{d-1}\,\mathrm{d}v}
{\sqrt{1-(-1)^{d-1}v^{2(d-1)}}}.
\label{S:eq:Fdonstraighttimelikeray}
\end{equation}
The behavior of this continuation depends qualitatively on the parity of
the boundary dimension $d$.
For odd $d$, Eq.~\eqref{S:eq:Fdonstraighttimelikeray} reduces to
\begin{equation}
F_d(-iu)
=
(-i)^d
\int_0^u
\frac{v^{d-1}\,\mathrm{d}v}
{\sqrt{1-v^{2(d-1)}}}.
\end{equation}
The square root vanishes at $u=1$, so the straight ray reaches the branch
point $\zeta=-i$. More generally, the branch points of the extremal
equation are
\begin{equation}
\zeta_k=e^{i\pi k/(d-1)},
\qquad
k=0,\ldots,2d-3,\nn
\end{equation}
and $\zeta=-i$ belongs to this set precisely when $d$ is odd.
Reaching this branch point, however, does not connect the two boundary
branches. On the sheet continuously connected to the origin,
\begin{equation}
F_d(-i)=(-i)^d c_d\in i\mathbb{R}.
\label{S:eq:Fdatminusi}
\end{equation}
The branches originating from the two boundary endpoints are
\begin{equation}
\lambda_{+}=c_d-F_d(\zeta),
\qquad
\lambda_{-}=-c_d+F_d(\zeta).
\label{S:eq:twoboundarybranches}
\end{equation}
They would meet at $\zeta=-i$ only if $F_d(-i)=c_d$, which is not
satisfied. Thus, although the straight ray reaches a branch point, it does
not by itself define a connected extremal surface joining the two boundary
components. Furthermore, in the $d=3$ example studied below, the complex
family metric associated with this continuation violates the KSW condition.
For even $d$, by contrast,
\begin{equation}
F_d(-iu)
=
(-i)^d
\int_0^u
\frac{v^{d-1}\,\mathrm{d}v}
{\sqrt{1+v^{2(d-1)}}}
\in\mathbb{R}.
\label{S:eq:Fdevend}
\end{equation}
The denominator never vanishes for real $u>0$, and hence the straight
ray encounters no branch point and can be continued to arbitrarily large
$u$. On the branch
\begin{equation}
\lambda(u)=c_d-F_d(-iu),
\label{S:eq:lambdaevenray}
\end{equation}
both $\lambda(u)$ and
\begin{equation}
z=\frac{i\rho_t}{c_d}\zeta
=\frac{\rho_t}{c_d}u
\end{equation}
are real. The corresponding embedding coordinates $t$, $x$, and $z$
are therefore all real. Wherever the coordinate map is nondegenerate, the
induced family metric is simply a real Lorentzian representation of
Poincaré AdS and consequently saturates the KSW bound.
In the $d=4$ example, the coordinate map remains nondegenerate for all
finite $u>0$. The straight ray can therefore be extended continuously to
$u\to\infty$, for which $z\to\infty$. It approaches the Poincaré
horizon rather than a second asymptotic boundary and hence still does not
provide a complete connected timelike extremal surface.
The role of the branch-point analysis is therefore to distinguish the fate
of the same natural straight continuation in odd and even dimensions. For
odd $d$, the ray terminates at a branch point without joining the two
boundary branches and, at least for $d=3$, leads to a violation of the KSW
condition. For even $d$, the ray encounters no branch point and gives a
KSW-allowable real Lorentzian branch, but it extends toward the Poincaré
horizon rather than returning to the second boundary. In neither case does
the straight continuation alone determine the complete gravitational
saddle.

In the following sections, we examine the admissible choices of $\lambda$ and the corresponding KSW conditions in greater detail for the AdS$_4$ and AdS$_5$ examples using a perturbative analysis. As we will see, the two dimensions exhibit qualitatively different behaviors.

\subsection{The straight real branch in $\AdS_5$}

For $d=4$, the extremal-surface equation is
\begin{align}
&c_4-\lambda
=
F_4(\zeta),
\qquad
F_4(\zeta)
=
\int_0^\zeta
\frac{s^3\,\mathrm{d}s}{\sqrt{1-s^6}}=
\frac{\zeta^4}{4}
{}_2F_1\left(
\frac{1}{2},\frac{2}{3};\frac{5}{3};\zeta^6
\right),
\label{S:eq:F4full}
\\
&c_4
=
\frac{1}{6} \mathrm{B}\left(\frac{2}{3},\frac{1}{2}\right)
=
\frac{\sqrt{\pi}\,\Gamma(2/3)}
{\Gamma(1/6)}.
\end{align}
Near the conformal boundary,
\begin{equation}
F_4(\zeta)
=
\frac{\zeta^4}{4}
+
\frac{\zeta^{10}}{20}
+
\frac{3\zeta^{16}}{128}
+
O(\zeta^{22}).
\label{S:eq:F4series}
\end{equation}
A general analytic contour compatible with the leading real-cutoff
condition $\zeta=-iu+O(u^2)$ may be parametrized as
\begin{align}
\lambda(u)
&=
c_4-\frac{u^4}{4}
+\alpha_1u^5+\alpha_2u^6+\alpha_3u^7
+O(u^8),
\label{S:eq:ads5lambdaexp}
\\
\zeta(u)
&=
-iu+i\alpha_1u^2
+i\left(
\alpha_2+\frac{3}{2}\alpha_1^2
\right)u^3
\nonumber\\
&\quad
+i\left(
\alpha_3+3\alpha_1\alpha_2
+\frac{7}{2}\alpha_1^3
\right)u^4
+O(u^5).
\label{S:eq:ads5zetaexp}
\end{align}
Here the coefficients $\alpha_n$ are generally complex. Equations
\eqref{S:eq:ads5lambdaexp} and
\eqref{S:eq:ads5zetaexp} follow by formally inverting
Eq.~\eqref{S:eq:F4series}; the $\zeta^{10}$ term first contributes at
higher order than those displayed above.

The perturbative expansion makes the distinction between the leading real-cutoff condition and an exactly real radial branch explicit. The leading condition $\zeta=-iu+O(u^2)$ does not constrain the coefficients $\alpha_n$ to be real. If one further requires $z(u)$ to remain real and positive in a neighborhood of the boundary, then $\zeta(u)$ must remain purely imaginary. Equation~\eqref{S:eq:ads5zetaexp} consequently implies, order by order,
\begin{equation}
\operatorname{Im}\alpha_n=0,
\qquad n\geq1.
\end{equation}
Indeed, once $\alpha_1,\ldots,\alpha_{n-1}$ are real, the coefficient at the next order is $\alpha_n$ plus a real polynomial in the preceding coefficients, so its reality requires $\alpha_n$ to be real. The remaining real coefficients correspond to real reparametrizations of the same negative-imaginary ray.
This argument establishes the uniqueness of the locally real radial branch, modulo real reparametrizations. It does not by itself prove that KSW allowability excludes all genuinely complex deformations of this branch. What follows is therefore an existence result: the real branch is exactly Lorentzian and saturates the KSW bound.

Along the negative-imaginary ray, write
\begin{equation}
\zeta=-iq,
\qquad q\in\mathbb{R}_{\geq0}.
\end{equation}
On the square-root branch continuously connected to the origin,
\begin{align}
F_4(-iq)
&=
\int_0^{-iq}
\frac{s^3\,\mathrm{d}s}{\sqrt{1-s^6}}
\nonumber\\
&=
\int_0^q
\frac{v^3\,\mathrm{d}v}{\sqrt{1+v^6}}
\equiv H_4(q)
\in\mathbb{R}.
\end{align}
Therefore,
\begin{equation}
\lambda(q)=c_4-H_4(q)
\label{S:eq:lambdaH4exact}
\end{equation}
is real for every real $q\geq0$. This proves, to all orders, the
existence of an exactly real timelike branch. Conversely, any real
analytic monotonic parametrization
\begin{equation}
q=q(u),
\qquad
q(0)=0,
\qquad
q'(0)=1,
\end{equation}
generates a formal expansion of the form
\eqref{S:eq:ads5lambdaexp}--\eqref{S:eq:ads5zetaexp} with real
coefficients $\alpha_n$. The real coefficients therefore describe
reparametrizations of the same negative-imaginary ray rather than
distinct complex contours. We may consequently choose
\begin{equation}
q(u)=u
\end{equation}
without loss of generality on this real branch, giving
\begin{equation}
\zeta=-iu,
\qquad
\lambda(u)=c_4-H_4(u).
\end{equation}
With $t_0=\rho_t\cosh\eta$ and $x_0=\rho_t\sinh\eta$,
define
\begin{equation}
T=\frac{\rho_t\lambda(u)}{c_4},
\qquad
Z=\frac{\rho_t q(u)}{c_4}.
\end{equation}
The bulk embedding becomes
\begin{equation}
t=T\cosh\eta,
\qquad
x=T\sinh\eta,
\qquad
z=Z,
\end{equation}
and the family metric is exactly
\begin{equation}
\dd s^2
=
\frac{L_{\text{AdS}}^2}{Z^2}
\left(
-\dd T^2+\dd Z^2
+T^2\dd\eta^2
+\dd y_1^2+\dd y_2^2
\right).
\label{S:eq:ads5realmetric}
\end{equation}
Thus the straight branch is not merely asymptotically real: wherever
the coordinate map is regular, it is an exact real Lorentzian
presentation of Poincaré $\AdS_5$. In a real Cartesian frame the
metric has one negative and four positive eigenvalues, and hence
\begin{equation}
\Theta[g]
=
\sum_A
\left|\arg\Lambda_A\right|
=
\pi.
\end{equation}
The KSW bound is therefore saturated exactly along this branch.
For completeness, the Jacobian of the radial transformation
$(\rho_t,u)\mapsto(T,Z)$ is
\begin{equation}
\det
\frac{\partial(T,Z)}
{\partial(\rho_t,u)}
=
\frac{\rho_t}{c_4^2}
\left(
\lambda q'-q\lambda'
\right).
\end{equation}
Using $\lambda=c_4-H_4(q)$, one obtains
\begin{equation}
\lambda q'-q\lambda'
=
q'G_4(q),
\qquad
G_4(q)
:=
c_4-H_4(q)+qH_4'(q).
\end{equation}
Since
\begin{equation}
G_4(0)=c_4>0,
\qquad
G_4'(q)
=
qH_4''(q)
=
\frac{3q^3}{(1+q^6)^{3/2}}>0,
\end{equation}
the radial map is nondegenerate for every $q>0$ whenever $q'>0$.
The full boost-adapted chart nevertheless degenerates when
$\lambda=0$, or equivalently $T=0$, because the Milne coordinate
$\eta$ collapses there. Since $H_4(q)$ is strictly increasing and
unbounded, this occurs once at a finite value of $q$. This is only a
coordinate caustic of the extremal-surface foliation: the original
Poincaré metric remains regular and real Lorentzian across it.
Finally, the negative-imaginary ray contains no branch point, since $1-(-iq)^6=1+q^6>0$ for all real $q$. The straight branch can therefore be continued to
arbitrarily large $q$, with
\begin{equation}
Z=\frac{\rho_t q}{c_4}\longrightarrow\infty.
\end{equation}
Within the Poincaré patch it approaches the Poincaré horizon rather
than returning to the second asymptotic boundary. Whether this branch
admits a global continuation that satisfies both timelike boundary
conditions is a separate question and is not determined by the local
or straight-ray analysis.
The result established here is therefore an existence statement: in
$\AdS_5$, the timelike real-cutoff condition admits an exact real
Lorentzian branch that saturates the KSW bound. It does not by itself
exclude additional complex KSW-allowable contours or uniquely determine
the complete gravitational saddle.

The straight real branch reaches the Poincar'e horizon as $u\to\infty$, but it does not return to the asymptotic boundary or connect to the second boundary branch. A complete surface satisfying both boundary conditions must therefore leave the real Lorentzian slice and follow a genuinely complex contour in the interior. Determining this global complex extremal surface, together with its KSW properties, lies beyond the scope of the present analysis.

\section{The $\AdS_4$ strip obstruction}
\label{sec:ads4strip}

For AdS$_4$ we have
\begin{equation}
 c_3-\lambda=\int_0^\zeta\frac{s^2\dd s}{\sqrt{1-s^4}},
 \qquad
 c_3=\int_0^1\frac{s^2\dd s}{\sqrt{1-s^4}}.
 \label{S:eq:ads4stripequation}
\end{equation}
As shown in Section~\ref{section_metric_strip}, setting $d=3$ in Eqs.~\eqref{S:eq:general-timelike-strip-metric} and \eqref{S:eq:general-spacelike-strip-metric} reproduces the adapted metrics for the timelike and spacelike strips in AdS$_4$, respectively.

\subsection{Spacelike branch}

We first identify the exact real spacelike branch. The two branches of the connected spacelike extremal surface are
\begin{equation}
\lambda_{+}(\zeta)
=
c_3-F_3(\zeta),
\qquad
\lambda_{-}(\zeta)
=
-c_3+F_3(\zeta),
\qquad
0\leq\zeta\leq1.
\label{S:eq:spaceliketwobranches}
\end{equation}
Since $F_3(\zeta)$ is real and monotonically increasing on this
interval, both $\lambda_\pm$ and $\zeta$ are real. The two branches
meet at
\begin{equation}
\zeta=1,
\qquad
\lambda_{+}=\lambda_{-}=0.
\label{S:eq:spaceliketurningpoint}
\end{equation}
Using $t_0=\rho_s\sinh\eta$ and $x_0=\rho_s\cosh\eta$,
introduce
\begin{equation}
X=\frac{\rho_s\lambda}{c_3},
\qquad
Z=\frac{\rho_s\zeta}{c_3}.
\label{S:eq:spacelikeXZ}
\end{equation}
The bulk embedding then becomes
\begin{equation}
t=X\sinh\eta,
\qquad
x=X\cosh\eta,
\qquad
z=Z.
\label{S:eq:spacelikeembeddingXZ}
\end{equation}
Consequently, the family metric takes the exact form
\begin{equation}
\dd s^2
=
\frac{L_{\text{AdS}}^2}{Z^2}
\left(
\dd Z^2+\dd X^2
-X^2\dd\eta^2+\dd y^2
\right).
\label{S:eq:spacelikeexactmetric}
\end{equation}
This is simply Lorentzian Poincaré $\AdS_4$ written in coordinates
adapted to the spacelike extremal surfaces. The apparent degeneration at
$X=0$ is only the usual degeneration of the boost coordinate
$\eta$; the bulk geometry itself is regular. In a real diagonal frame,
the metric has one negative and three positive eigenvalues, and therefore
saturates the KSW bound,
\begin{equation}
\sum_A\left|\arg\Lambda_A\right|=\pi.
\end{equation}
Thus, the existence of a real Lorentzian spacelike branch is an exact,
nonperturbative result.
We next examine whether a regular complex deformation of this branch can
remain KSW-compatible near the conformal boundary. A general analytic
contour satisfying the leading positive-real-cutoff condition may be
written as
\begin{align}
\lambda(u)
&=
c_3-\frac{u^3}{3}
+\alpha_1u^4+\alpha_2u^5+\alpha_3u^6
+O(u^7),
\label{S:eq:spacelikelambdaexp}
\\
\zeta(u)
&=
u-\alpha_1u^2
-\left(\alpha_2+\alpha_1^2\right)u^3
\nonumber\\
&\quad
-\left(
\alpha_3+2\alpha_1\alpha_2
+\frac{5}{3}\alpha_1^3
\right)u^4
+O(u^5),
\end{align}
where
\begin{equation}
\alpha_n=\alpha_{nR}+i\alpha_{nI}.
\end{equation}
The leading cutoff condition fixes the leading term
$\zeta=u+O(u^2)$, but does not by itself require the coefficients
$\alpha_n$ to be real.
A necessary KSW condition (\ref{S:eq:matrixnecessary}) is the positive semidefiniteness of $M^{ab}:=\operatorname{Re}\left(\sqrt{g}\,g^{ab}\right)$.
The first nonvanishing terms are
\begin{align}
M^{uu}
&=
-\frac{4c_3^3\alpha_{1I}}
{u\rho_s^2}
+O(u^0),
\nonumber\\
M^{u\rho_s}
&=
\frac{c_3\alpha_{1I}}{\rho_s}
+O(u),
\nonumber\\
M^{\rho_s\rho_s}
&=
c_3K_2+O(u),
\nonumber\\
M^{\eta\eta}
&=
-\frac{c_3K_2}{\rho_s^2}
+O(u),
\nonumber\\
M^{yy}
&=
c_3K_2+O(u),
\end{align}
where $K_2:=4\alpha_{1I}\alpha_{1R}+\alpha_{2I}$.
Positivity of $M^{uu}$ first requires
\begin{equation}
\alpha_{1I}\leq0.
\end{equation}
Since $M^{\eta\eta}$ and $M^{yy}$ contain $K_2$ with opposite
signs, semidefiniteness requires
\begin{equation}
K_2=0.
\end{equation}
The next nonvanishing contribution to the
$(u,\rho_s)$ principal minor is then negative unless
\begin{equation}
\alpha_{1I}=0.
\end{equation}
It follows from $K_2=0$ that
\begin{equation}
\alpha_{2I}=0.
\label{S:eq:spacelikeconstraints12}
\end{equation}
At the following order, the same positivity condition gives
\begin{equation}
\alpha_{3I}=0.
\end{equation}
The recursive structure can be stated more generally. Suppose that
$\alpha_1,\ldots,\alpha_{n-1}$ are real and that
$\alpha_n$ is the first coefficient with a nonzero imaginary part.
For $n\geq2$, the first terms depending on $\alpha_{nI}$ appear as
\begin{align}
M^{\eta\eta}
&=
-\frac{c_3(n-1)}{\rho_s^2}
\alpha_{nI}u^{n-2}
+O(u^{n-1}),
\nonumber\\
M^{yy}
&=
c_3(n-1)
\alpha_{nI}u^{n-2}
+O(u^{n-1}).
\end{align}
The two diagonal entries have opposite signs unless
\begin{equation}
\alpha_{nI}=0.
\end{equation}
Together with the separate analysis of $\alpha_{1I}$, this gives,
by induction,
\begin{equation}
\alpha_n\in\mathbb{R},
\qquad
n\geq1.
\label{S:eq:allrealspacelike}
\end{equation}
The remaining real coefficients correspond only to real analytic
reparametrizations of the positive-real $\zeta$ axis. Hence, modulo a
real reparametrization, the regular analytic KSW-compatible contour near
the boundary is the standard real branch. Assuming analytic continuation
without introducing an additional nonanalytic segment, this local branch
extends uniquely along
\begin{equation}
\lambda=c_3-F_3(\zeta),
\qquad
0\leq\zeta\leq1,
\end{equation}
and reaches the real turning point
$(\lambda,\zeta)=(0,1)$. The two boundary branches then join to form the
ordinary real spacelike extremal surface.
We therefore obtain two complementary results. First, the standard
spacelike family gives an exact Lorentzian presentation of
$\AdS_4$ and saturates the KSW bound nonperturbatively. Second, within
the class of regular analytic contours connected to a positive real
cutoff, the perturbative KSW conditions eliminate all complex
deformations order by order and locally select this real branch, up to
real reparametrizations.

\subsection{Timelike branch and KSW obstruction}
For timelike separation, a general analytic contour compatible with the
positive-real-cutoff condition near the conformal boundary may be written as
\begin{align}
\lambda(u)
&=
c_3-\frac{i}{3}u^3
+\alpha_1u^4
+\alpha_2u^5
+\alpha_3u^6
+O(u^7),
\label{S:eq:timelikelambdaexp}
\\
\zeta(u)
&=
-iu+\alpha_1u^2
+\left(\alpha_2-i\alpha_1^2\right)u^3
\nonumber\\
&\quad
+\left(
\alpha_3-2i\alpha_1\alpha_2-\frac{5}{3}\alpha_1^3
\right)u^4
+O(u^5),
\label{S:eq:timelikezetaexp}
\end{align}
where $\alpha_n=\alpha_{nR}+i\alpha_{nI}$.
The real-cutoff condition fixes the leading behavior
$\zeta=-iu+O(u^2)$, but does not by itself determine the subleading
coefficients $\alpha_n$.

In the basis $(u,\rho_t,\eta,y)$, the metric components required for the
KSW analysis are
\begin{align}
g_{uu}
={}&
\frac{1}{u^2}
+\frac{2i\alpha_1}{u}
+5\alpha_1^2
+4i\alpha_2
\nonumber\\
&\quad
+2u\left(
-6i\alpha_1^3
+7\alpha_1\alpha_2
+3i\alpha_3
\right)
+O(u^2),
\label{S:eq:tguu}
\\
g_{u\rho}
={}&
\frac{1}{u\rho_t}
+\frac{i(c_3+\alpha_1)}{\rho_t}
+\frac{
u\left(
-2c_3\alpha_1
+3\alpha_1^2
+2i\alpha_2
\right)
}{\rho_t}
+O(u^2),
\label{S:eq:tgurho}
\\
g_{\rho\rho}
={}&
-\frac{c_3^2}{u^2\rho_t^2}
+\frac{2ic_3^2\alpha_1}{u\rho_t^2}
+\frac{
1+c_3^2\left(5\alpha_1^2+2i\alpha_2\right)
}{\rho_t^2}
\nonumber\\
&\quad
-\frac{2iu}{3\rho_t^2}
\left[
-c_3
+20c_3^2\alpha_1^3
+15ic_3^2\alpha_1\alpha_2
-3c_3^2\alpha_3
\right]
+O(u^2),
\label{S:eq:tgrhorho}
\\
g_{\eta\eta}
={}&
\frac{c_3^2}{u^2}
-\frac{2ic_3^2\alpha_1}{u}
-c_3^2\left(5\alpha_1^2+2i\alpha_2\right)
\nonumber\\
&\quad
+u\left[
-\frac{2ic_3}{3}
-\frac{2c_3^2}{3}
\left(
-20i\alpha_1^3
+15\alpha_1\alpha_2
+3i\alpha_3
\right)
\right]
+O(u^2),
\label{S:eq:tgetaeta}
\\
g_{yy}
={}&
\frac{c_3^2}{u^2\rho_t^2}
-\frac{2ic_3^2\alpha_1}{u\rho_t^2}
+\frac{
c_3^2\left(-5\alpha_1^2-2i\alpha_2\right)
}{\rho_t^2}
\nonumber\\
&\quad
-\frac{2uc_3^2}{3\rho_t^2}
\left(
-20i\alpha_1^3
+15\alpha_1\alpha_2
+3i\alpha_3
\right)
+O(u^2).
\label{S:eq:tgyyfull}
\end{align}
A necessary condition for KSW allowability is the positive
semidefiniteness of $M^{ab}
:=
\operatorname{Re}\left(\sqrt{g}\,g^{ab}\right)$.
Before imposing any constraints on the coefficients, its nonzero components
take the form
\begin{align}
M^{uu}
={}&
\frac{4c_3^3\alpha_{1R}}{u\rho_t^2}
+\frac{c_3^3}{\rho_t^2}
\left(
32c_3^2\alpha_{1I}\alpha_{1R}
+5c_3^2\alpha_{2R}
-2\alpha_{1R}
\right)
+O(u),
\label{S:eq:MuuExplicit}
\\
M^{u\rho}
={}&
\frac{c_3(\alpha_{1R}-c_3)}{\rho_t}
+\frac{
uc_3\left(
4\alpha_{1I}\alpha_{1R}
+\alpha_{2R}
\right)
}{\rho_t}
+O(u^2),
\label{S:eq:MurhoExplicit}
\\
M^{\rho\rho}
={}&
c_3\left(
4\alpha_{1I}\alpha_{1R}
+\alpha_{2R}
\right)
\nonumber\\
&\quad
+u\left[
c_3\left(
28\alpha_{1I}^2\alpha_{1R}
+8\alpha_{1I}\alpha_{2R}
-\frac{28}{3}\alpha_{1R}^3
+8\alpha_{1R}\alpha_{2I}
+2\alpha_{3R}
\right)
-1
\right]
+O(u^2),
\label{S:eq:MrhorhoExplicit}
\\
M^{\eta\eta}
={}&
-\frac{
c_3\left(
4\alpha_{1I}\alpha_{1R}
+\alpha_{2R}
\right)
}{\rho_t^2}
\nonumber\\
&\quad
-\frac{u}{3\rho_t^2}
\left[
c_3\left(
84\alpha_{1I}^2\alpha_{1R}
+24\alpha_{1I}\alpha_{2R}
-28\alpha_{1R}^3
+24\alpha_{1R}\alpha_{2I}
+6\alpha_{3R}
\right)
+3
\right]
+O(u^2),
\label{S:eq:MetaetaExplicit}
\\
M^{yy}
={}&
-c_3\left(
4\alpha_{1I}\alpha_{1R}
+\alpha_{2R}
\right)
\nonumber\\
&\quad
-\frac{u}{3}
\left[
c_3\left(
84\alpha_{1I}^2\alpha_{1R}
+24\alpha_{1I}\alpha_{2R}
-28\alpha_{1R}^3
+24\alpha_{1R}\alpha_{2I}
+6\alpha_{3R}
\right)
+1
\right]
+O(u^2).
\label{S:eq:MyyExplicit}
\end{align}
The leading $u^{-1}$ term in $M^{uu}$ first requires
\begin{equation}
\alpha_{1R}\geq0.
\label{S:eq:timelikealpha1positive}
\end{equation}
The constant terms in $M^{\rho\rho}$, $M^{\eta\eta}$, and $M^{yy}$
contain the same combination with opposite signs. Their simultaneous
nonnegativity therefore requires
\begin{equation}
4\alpha_{1I}\alpha_{1R}+\alpha_{2R}=0.
\label{S:eq:timelikecombinationzero}
\end{equation}
After imposing Eq.~\eqref{S:eq:timelikecombinationzero}, positivity of the
Schur complement of the $(u,\rho_t)$ block sharpens
Eq.~\eqref{S:eq:timelikealpha1positive} to
\begin{equation}
\alpha_{1R}=c_3.
\label{S:eq:timelikealpha1constraint}
\end{equation}
Equation~\eqref{S:eq:timelikecombinationzero} then gives
\begin{equation}
\alpha_{2R}=-4c_3\alpha_{1I}.
\label{S:eq:timelikealpha2constraint}
\end{equation}
Thus the first two nontrivial orders yield the necessary conditions
\begin{equation}
\alpha_{1R}=c_3,
\qquad
\alpha_{2R}=-4c_3\alpha_{1I}.
\label{S:eq:timelikenecessaryconstraints}
\end{equation}
These relations remove the leading negative directions of $M^{ab}$.
They are only necessary conditions and do not establish full KSW
allowability. In particular, the imaginary part $\alpha_{1I}$, as well
as $\alpha_{2I}$ and the higher coefficients, remains unconstrained at
this stage.

Substituting Eq.~\eqref{S:eq:timelikenecessaryconstraints} into the next
nonvanishing terms of $M^{ab}$, one finds incompatible positivity
requirements among the $\rho_t$, $\eta$, and $y$ directions. No
choice of $\alpha_{3R}$ and $\alpha_{3I}$ can make all principal minors
positive semidefinite at this order. Hence the necessary scalar KSW
condition already fails in a sufficiently small neighborhood of the
conformal boundary.
The calculation above is expressed in the original non-diagonal family
coordinates. In the following subsection, we construct a local real frame
in which the metric is diagonal to the required order. The same obstruction
then appears invariantly as a positive excess of the total KSW phase,
\begin{equation}
\sum_A\left|\arg\Lambda_A\right|>\pi,
\end{equation}
as displayed explicitly in next section.

\subsection{Local real diagonalization}
Define a real radial coordinate $r$ by
\begin{equation}
r = \rho_t u -\rho_t \alpha_{1I} u^2 +\rho_t \left( 2c_3^2-\alpha_{1I}^2-\alpha_{2I} \right)u^3 +O(u^4).
\label{S:eq:rcoordinate}
\end{equation}
Equivalently, a real one-form of the form
\begin{equation}
\widehat V = \mathrm{d} u+\mu(u)\frac{\mathrm{d}\rho_t}{\rho_t}
\end{equation}
admits an integrating factor such that $\widehat V \propto \frac{\mathrm{d} r}{r}$.
After imposing Eq.~\eqref{S:eq:timelikenecessaryconstraints}, the metric in the real coordinates $(r, \rho_t, \eta, y)$ is diagonal through the universal orders relevant to the phase sum:
\begin{align}
g_{rr}
&=
\frac{1}{r^2}
+\frac{2ic_3}{r\rho_t}
-\frac{3c_3^2}{\rho_t^2}
+O(r),
\label{S:eq:grr} \\
g_{\rho\rho}
&=
-\frac{c_3^2}{r^2}
+\frac{2ic_3^3}{r\rho_t}
+\frac{c_3^4}{\rho_t^2}
+O(r),
\label{S:eq:grhorho} \\
g_{\eta\eta}
&=
\frac{c_3^2\rho_t^2}{r^2}
-\frac{2ic_3^3\rho_t}{r}
-c_3^4
+O(r),
\label{S:eq:getaeta} \\
g_{yy}
&=
\frac{c_3^2}{r^2}
-\frac{2ic_3^3}{r\rho_t}
-\frac{c_3^4}{\rho_t^2}
+O(r).
\label{S:eq:gyy}
\end{align}
The coefficients displayed in Eqs.~\eqref{S:eq:grr}--\eqref{S:eq:gyy} are independent of the remaining unconstrained imaginary data at this order. The $O(r^{-1})$ terms are therefore not removable by tuning the next coefficient of the contour.

\subsection{Full phase sum}

Set
\begin{equation}
 \kappa=\frac{c_3r}{\rho_t}>0.
\end{equation}
Factoring the leading magnitudes from Eqs.~\eqref{S:eq:grr}--\eqref{S:eq:gyy}, their phases are
\begin{align}
 |\Arg g_{rr}|&=2\kappa+O(\kappa^2),
 \nonumber\\
 |\Arg g_{\rho\rho}|&=\pi-2\kappa+O(\kappa^2),
 \nonumber\\
 |\Arg g_{\eta\eta}|&=2\kappa+O(\kappa^2),
 \nonumber\\
 |\Arg g_{yy}|&=2\kappa+O(\kappa^2).
\end{align}
Consequently
\begin{equation}
 \Theta[g]=\pi+4\frac{c_3r}{\rho_t}+O(r^2)>\pi,
 \qquad(r>0\ \text{sufficiently small}).
 \label{S:eq:stripviolation}
\end{equation}
Since $z\propto r\propto u$ near the boundary, this is the $O(z)$ violation stated in the main text. The timelike strip approaches the Lorentzian boundary of the KSW cone from the nonallowable side. This differs qualitatively from the locally-$\AdS_3$, $\dS_3$, and hyperbolic contours, which remain exactly on $\Theta=\pi$ segment by segment.

\subsection{A diagonalizability condition}

In the previous section, we did not impose any additional constraints on the contour $\lambda$. The examples of $\text{AdS}_3$ and $\text{AdS}_{d+1}$ with even boundary dimension $d$ suggest that $\lambda$ may be chosen to be real-valued, at least along the relevant local branch. In this subsection, we impose an additional condition on the contour $\lambda$. As shown below, this condition ensures that the metric in Eq.~\eqref{S:eq:general-timelike-strip-metric} can be diagonalized by a real coordinate transformation. Subject to this constraint, the contour $\lambda$ can be determined either perturbatively near the asymptotic boundary or numerically in the bulk. Nevertheless, even after imposing this stronger condition, we find that the resulting $\text{AdS}_4$ geometry still violates the KSW criterion.

For completeness, a sufficient condition for diagonalizing the $(u,\rho_t)$ block by a real coordinate transformation is
\begin{equation}
 \Ima(\lambda^2+\zeta^2)=0.
 \label{S:eq:strongcondition}
\end{equation}
Indeed, with $R(u)=\lambda^2+\zeta^2$, the ratio of the off-diagonal and radial entries satisfies
\begin{equation}
 \frac{g_{\rho u}}{g_{\rho\rho}}=\frac{\rho_tR'}{2R}.
\end{equation}
If $R$ is real, then
\begin{equation}
 \dd V_t=\frac{\dd\rho_t}{\rho_t}+\frac12\frac{R'}{R}\dd u
 =\dd\log\!\left(\rho_t\sqrt{R}\right)
\end{equation}
provides a real diagonal coordinate.  Completing the square gives the exact diagonal form
\begin{align}
 \frac{\dd s^2}{L_{\text{AdS}}^2}={}&
 \frac{\zeta^2+\lambda^2}{\zeta^2}\dd V_t^2
 -\frac{\lambda^2}{\zeta^2}\dd\eta^2
 -\frac{c_3^2e^{-2V_t}(\zeta^2+\lambda^2)}{\zeta^2}\dd y^2
 \nonumber\\
 &+\frac{(\lambda\sqrt{1-\zeta^4}+\zeta^3)^2}
 {\zeta^6(\zeta^2+\lambda^2)}\lambda'^2\dd u^2.
\end{align}
For the spacelike branch, the transverse $y$ term has the opposite sign, while the same real coordinate construction applies.  Solving Eq.~\eqref{S:eq:strongcondition} perturbatively on the timelike branch gives, for example,
\begin{align}
 \lambda(u)&=c_3-\frac{i}{3}u^3-\frac{c_3}{3}u^4
 +\frac{ic_3^2}{9}u^5+\frac{10c_3^3}{81}u^6+O(u^7),
 \nonumber\\
 \zeta(u)&=-iu-\frac{c_3}{3}u^2+\frac{c_3^3}{9}u^4+O(u^5).
\end{align}

Substituting these expansions into the diagonal family metric gives
\begin{align}
ds^2
={}&
\frac{1}{u^2}
\left(
-c_3^2 \, dV_t^2
+c_3^2 \, d\eta^2
+c_3^4 e^{-2V_t} \, dy^2
+du^2
\right)
\nonumber \\
&\quad
-\frac{2ic_3}{3u}
\left(
c_3^2 \, dV_t^2
-c_3^2 \, d\eta^2
-c_3^4 e^{-2V_t} \, dy^2
+du^2
\right)
+O(u^0).
\label{Perturbation_AdS4}
\end{align}
Several features of Eq.~\eqref{Perturbation_AdS4} are worth emphasizing. First, the metric is manifestly complex, even though it has been diagonalized in a real coordinate system. Second, it satisfies the Einstein equations order by order in the near-boundary expansion. This is expected because the metric is obtained from complexified Poincaré AdS by a coordinate transformation and therefore remains locally an Einstein metric.

The leading term in Eq.~\eqref{Perturbation_AdS4} has the standard asymptotically AdS form and defines a real Lorentzian boundary metric. However, the first subleading contribution appears at order $O(u^{-1})$ and is purely imaginary. Unlike the usual Fefferman--Graham expansion of a real asymptotically AdS metric, this term cannot be removed by any real coordinate transformation that preserves the real integration cycle. It therefore represents an intrinsic feature of the complex real-dimensional section selected by the extremal-surface contour.

The condition~\eqref{S:eq:strongcondition} guarantees real diagonalizability, but it does not guarantee KSW allowability. Since the metric~\eqref{Perturbation_AdS4} is already diagonal in a real basis, the full KSW criterion can be tested directly from the phases of its diagonal eigenvalues.

Define
\begin{equation}
\delta\phi := \arctan\left(\frac{2c_3u}{3}\right).
\label{S:eq:deltaphi}
\end{equation}
To the first nontrivial order, the four diagonal components have phase magnitudes
\begin{align}
\left|\operatorname{Arg}g_{V_tV_t}\right|
&=
\pi-\delta\phi+O(u^2),
\nonumber \\
\left|\operatorname{Arg}g_{\eta\eta}\right|
&=
\delta\phi+O(u^2),
\nonumber \\
\left|\operatorname{Arg}g_{yy}\right|
&=
\delta\phi+O(u^2),
\nonumber \\
\left|\operatorname{Arg}g_{uu}\right|
&=
\delta\phi+O(u^2).
\end{align}
Consequently, the KSW phase sum is
\begin{equation}
\sum_i \left| \operatorname{Arg}\lambda_i \right| = \pi+2\delta\phi+O(u^2),
\qquad
\delta\phi = \arctan\left(\frac{2c_3u}{3}\right).
\label{S:eq:stripviolation_diag}
\end{equation}
For $u>0$, one has $\delta\phi>0$, and hence
\begin{equation}
\sum_i \left| \operatorname{Arg}\lambda_i \right| > \pi.
\end{equation}
The KSW criterion is therefore violated immediately away from the conformal boundary. Since
\begin{equation}
\delta\phi = \frac{2c_3}{3}u+O(u^3),
\end{equation}
the violation first appears at order $O(u)$, in agreement with Eq.\eqref{S:eq:stripviolation}

\section{$\text{AdS}_3$ perturbative consistency check}
\label{sec:ads3pert}
To verify the reliability of the perturbative method developed for higher-dimensional examples, we now apply the same near-boundary analysis to $\text{AdS}_3$. As shown below, the perturbative calculation reproduces the conclusion obtained from the exact, nonperturbative KSW analysis.

Write the geodesic as
\begin{equation}
t=t_0\lambda,
\qquad
x=x_0\lambda,
\qquad
z=\sqrt{x_0^2-t_0^2}\,\zeta(\lambda),
\qquad
\zeta(\lambda)=\sqrt{1-\lambda^2}.
\end{equation}
After introducing $\rho_s$ or $\rho_t$, together with the corresponding boost rapidity, the family metric becomes
\begin{equation}
\frac{\mathrm{d}s^2}{L^2} = \frac{\lambda'^2}{(\lambda^2-1)^2}\mathrm{d}u^2 -\frac{1}{\rho^2(\lambda^2-1)}\mathrm{d}\rho^2 +\frac{\lambda^2}{\lambda^2-1}\mathrm{d}\eta^2.
\label{S:eq:ads3pertmetric}
\end{equation}
Near $\lambda=1$, consider the expansion
\begin{equation}
\lambda(u) = 1-\frac{e^{i\theta}}{2}u^2 +\sum_{n\geq1}\alpha_nu^{n+2}.
\label{S:eq:ads3lambdaexp}
\end{equation}
For spacelike separation,
\begin{equation}
z\sim\rho_su e^{i\theta/2}>0
\end{equation}
fixes $\theta=0$. For timelike separation,
\begin{equation}
z\sim i\rho_tu e^{i\theta/2}>0
\end{equation}
fixes $\theta=-\pi$.

Expanding Eq.~\eqref{S:eq:ads3pertmetric}, one obtains
\begin{align}
\frac{\mathrm{d}s^2}{L^2}
={}& \left[ \frac{1}{u^2} -\frac{2\alpha_1e^{-i\theta}}{u} +A_u^{(0)} +A_u^{(1)}u +O(u^2) \right]\mathrm{d}u^2
\nonumber \\
& + \left[ \frac{e^{-i\theta}}{\rho^2u^2} +\frac{2\alpha_1e^{-2i\theta}}{\rho^2u} +A_\rho^{(0)} +A_\rho^{(1)}u +O(u^2) \right]\mathrm{d}\rho^2
\nonumber \\
& + \left[ -\frac{e^{-i\theta}}{u^2} -\frac{2\alpha_1e^{-2i\theta}}{u} +A_\eta^{(0)} +A_\eta^{(1)}u +O(u^2) \right]\mathrm{d}\eta^2.
\label{S:eq:ads3pertexpanded}
\end{align}
The coefficients appearing at orders $u^0$ and $u^1$ are
\begin{align}
A_u^{(0)} &= \frac{1}{2}e^{-2i\theta} \left( -6\alpha_1^2 -8\alpha_2e^{i\theta} +e^{3i\theta} \right), \nonumber \\
A_\rho^{(0)} &= \frac{e^{-3i\theta}}{4\rho^2} \left( 16\alpha_1^2 +8\alpha_2e^{i\theta} +e^{3i\theta} \right), \nonumber \\
A_\eta^{(0)} &= -\frac{1}{4}e^{-3i\theta} \left( 16\alpha_1^2 +8\alpha_2e^{i\theta} +e^{3i\theta} \right) +1, \nonumber \\
A_u^{(1)} &= -2e^{-3i\theta} \left( 2\alpha_1^3 +4\alpha_1\alpha_2e^{i\theta} +\alpha_1e^{3i\theta} +3\alpha_3e^{2i\theta} \right), \nonumber \\
A_\rho^{(1)} &= \frac{2e^{-4i\theta}}{\rho^2} \left( 4\alpha_1^3 +4\alpha_1\alpha_2e^{i\theta} +\alpha_3e^{2i\theta} \right), \nonumber \\
A_\eta^{(1)} &= -2e^{-4i\theta} \left( 4\alpha_1^3 +4\alpha_1\alpha_2e^{i\theta} +\alpha_3e^{2i\theta} \right).
\label{S:eq:ads3Acoeffs}
\end{align}
Writing $\alpha_n=\alpha_{nR}+i\alpha_{nI}$, the KSW condition at order $u^{-1}$ forces $\alpha_{1I}=0$. At the next order, one finds $\alpha_{2I}=0$. The same structure persists recursively, implying $\alpha_{nI}=0$ for all $n$. Equivalently,
\begin{equation}
\alpha_n\in\mathbb{R} \qquad \text{for all }n.
\end{equation}
The boundary branch is therefore described by a real geodesic parameter on the appropriate side of the branch point: $\lambda<1$ for spacelike separation, whereas $\lambda>1$ for timelike separation.

Indeed, for the spacelike branch, $\theta=0$ gives $\lambda = 1-\frac{u^2}{2}+O(u^3)<1$, while for the timelike branch, $\theta=-\pi$ gives $\lambda = 1+\frac{u^2}{2}+O(u^3)>1$. In particular, the intrinsically complex $O(u^{-1})$ deformation that is unavoidable for the timelike strip in $\text{AdS}_4$ is absent in $\text{AdS}_3$. The perturbative analysis therefore agrees with the exact $\text{AdS}_3$ contour theorem derived above, providing a nontrivial consistency check of the higher-dimensional perturbative method.

\section{Summary of the results}

The calculations establish three distinct facts.

First, the single complex family metric shared by Poincar\'e $\AdS_3$, global $\AdS_3$, and BTZ obeys one exact contour-selection theorem, and the $\AdS_4$ hyperbolic family inherits the same AdS contour within the stated regular class. In $\dS_3$, the opposite endpoint phases and the full KSW phase cone likewise fix the reflected three-piece contour. The apparent smooth alternatives satisfy only a weaker ratio inequality and fail the complete phase test.

Second, for the $\AdS_4$ timelike strip, the real-cutoff boundary condition and the extremal equation fix a local complex branch. Necessary KSW positivity constrains its first free coefficients, but the complete phase sum in a real diagonal frame exceeds $\pi$ at linear order in the radial coordinate. This is a genuine local exclusion result.

Third, the even--odd distinction for planar strips is a theorem about the leading boundary phase and the location of branch points. It is not, without further calculation, a universal parity theorem for KSW admissibility or for the contributing gravitational saddle. KSW is a necessary local filter on a proposed real cycle; Picard--Lefschetz intersection numbers and replica boundary conditions are still required to establish that a surviving cycle contributes to the Lorentzian gravitational path integral.

Let us finally comment on the marginal allowability and the role of the regulator. The original KSW domain is open, with $\Theta<\pi$, whereas the contours selected above satisfy $\Theta=\pi$. This is the same marginality as a real Lorentzian metric. Operationally, the equations should be understood as the zero-regulator limit of a contour displaced by the boundary $i\epsilon$ prescription into the allowable domain. The limiting analysis remains discriminating: generic deformations of $\cC_{\AdS}$ move some point of the induced metric to $\Theta>\pi$, while the selected image stays on the boundary everywhere. For the strip, Eq.~\eqref{eq:violation} has a positive excess at every sufficiently small but finite $z$. It therefore approaches the Lorentzian boundary from the nonallowable side, rather than merely saturating it.

This observation also separates two uses of complexification. A complex parametrization of a fixed real Lorentzian cycle is physically innocuous only when it can be undone by a real change of integration variables. By contrast, the timelike contour changes the real section of the complexified bulk, and the strip branch contains a complex $O(z^{-1})$ term in its FG-like expansion that cannot be removed by a real diffeomorphism. KSW is sensitive precisely to this choice of real section. It does not classify complex metrics modulo arbitrary holomorphic coordinate transformations, because such transformations can rotate the matter-field integration contour and alter convergence.


\begin{thebibliography}{99}
\bibitem{note1}
 Notice that the traces $\Tr$ and tr on the left- and right-hand sides are different. $\Tr$ denotes the trace over $\mathcal{H}_0\otimes \mathcal{H}_1$, while tr denotes the trace over $\mathcal{H}_0$.
\bibitem{note_parameter}

For the highly symmetric subregions considered in this work, the
nontrivial complexification can be reduced to the single coordinate
$\lambda$, while the coordinates $\sigma^A$ may be kept real. For more
general subregions without these symmetries, one may need to complexify
some or all of the $\sigma^A$ as well, leading to a multidimensional contour
problem that is considerably more involved.
\bibitem{note_marginal}
The strict KSW condition is $\Theta[g]<\pi$. In the following, we also use
its closure, $\Theta\leq\pi$, with equality understood as the
zero-$i\epsilon$ Lorentzian limit.
\end{thebibliography}
\end{document}